\documentclass[iop]{emulateapj}
\usepackage[latin1]{inputenc}
\usepackage[english]{babel}
\usepackage{amsmath}
\usepackage{amssymb}		
\usepackage{latexsym} 
\usepackage{epsfig}
\usepackage{subfigure}
\usepackage{natbib}
\usepackage{setspace}
\usepackage{graphicx}
\usepackage{longtable}
\newcommand{\msun}{{M_{\odot}}}
\newcommand{\mstar}{{M_{\ast}}}
\newcommand{\ser}{S\'ersic }

\shorttitle{On the robustness of $z=0-1$ galaxy size measurements}
\shortauthors{Mosleh et al.}

\begin{document}

\title{On The Robustness of $z=0-1$ Galaxy Size Measurements Through Model and Non-Parametric Fits}

\author{Moein Mosleh \altaffilmark{1}, Rik J. Williams \altaffilmark{2}, Marijn Franx \altaffilmark{1}}

\altaffiltext{1}{Leiden Observatory, Universiteit Leiden, 2300 RA Leiden, The Netherlands}\email{mosleh@strw.leidenuniv.nl}
\altaffiltext{2}{Carnegie Observatories, Pasadena, CA 91101, USA}

\begin{abstract}

We present the size-stellar mass relations of nearby ($z=0.01-0.02$) Sloan Digital Sky Survey galaxies, for samples selected by color, morphology, \ser index $n$, and specific star formation rate. Several commonly employed size measurement techniques are used, including single \ser fits, two-component \ser models, and a non-parametric method.  Through simple simulations, we show that the non-parametric and two-component \ser methods provide the most robust effective radius measurements, while those based on single \ser profiles are often overestimates, especially for massive red/early-type galaxies. Using our robust sizes, we show for all sub-samples that the mass-size relations are shallow at low stellar masses and steepen above $\sim3-4\times10^{10}\msun$.  The mass-size relations for galaxies classified as late-type, low-$n$, and star-forming are consistent with each other, while blue galaxies follow a somewhat steeper relation. The mass-size relations of early-type, high-$n$, red, and quiescent galaxies all agree with each other but are somewhat steeper at the high-mass end than previous results. To test potential systematics at high redshift, we artificially redshifted our sample (including surface brightness dimming and degraded resolution) to $z=1$ and re-fit the galaxies using single \ser profiles.  The sizes of these galaxies before and after redshifting are consistent and we conclude that systematic effects in sizes and the size-mass relation at $z\sim 1$ are negligible.  Interestingly, since the poorer physical resolution at high redshift washes out bright galaxy substructures, single-\ser fitting appears to provide more reliable and unbiased effective radius measurements at high $z$ than for nearby, well-resolved galaxies.\\

\end{abstract}

 
\section{Introduction}

Correlations among galaxy physical parameters such as stellar mass, luminosity, size, velocity dispersion, and their evolution with cosmic time are crucial for understanding the formation and evolution of galaxies and imposing constraints on theoretical models of their structural assembly. Morphological scaling relations such as the relation between size and surface brightness, the correlation between size and luminosity \citep{kormendy85},  and the relation between the effective radius and stellar mass \citep[hereafter S03]{shen2003}, vary for different types of galaxies. The differences between surface brightness profiles and sizes of galaxies are the products of the different physical processes governing their formation and evolution. Precise measurements of these galaxy properties at low and high redshifts thus provide strong constraints on models of galaxy formation and evolution.  

Among these relations is the observed correlation between half-light radius (size) and stellar mass, which is shown for the local universe (S03) and persists up to very high redshifts \citep[e.g.,][]{daddi2005, trujillo2006a, franx2008, buitrago2008, cimatti2008, vanderwel2008, williams2009, dutton2011, law2011, mosleh2011, mosleh2012}. These authors also pointed out that sizes of galaxies at fixed stellar mass decrease as  redshift increases, i.e., galaxies were smaller in the past.  For instance, massive quiescent galaxies at $z\sim2$ are about a factor of  $\sim6$ smaller than their counterparts at $z\sim0$ \citep[e.g.,][]{daddi2005,vandokkum2008}. Understanding the mechanism of the size evolution and how galaxies reach the mass-size relation at $z=0$ requires measuring these properties, especially sizes, very robustly in different redshift ranges.  

One of the main concerns is the accuracy of galaxy size determination at high redshifts. Galaxies at higher redshifts are at larger distances and therefore are dimmer and have smaller apparent angular sizes. The low surface brightness envelopes of galaxies could fade away due to cosmological dimming and could have lower signal to noise ratios (S/N), hence, potentially invoking systematics on the real size measurements. For example, the outer parts of early-type galaxies normally fade away gradually into the background sky noise and it is very hard to define precise edges for these types of galaxies.  Underestimating the sizes of these galaxies at high redshifts could have an effect on the inferred rate of size evolution \citep{mancini2010}. 

There are several possible approaches to test the compactness of galaxies at high redshifts. Recently, \cite{szomoru2010}  used deep observations with the Wide Field Camera 3 (WFC3) instrument on board the \textit{Hubble Space Telescope} (HST) to measure the size of a massive quiescent galaxy at $z\sim2$ based on a new approach (correcting the best-fit \ser profile of the galaxy with the residual of the fit) to confirm the compactness of this massive galaxy at this redshift.  

The other method to check the effects of cosmological redshift on the size/shape measurements is to artificially transform nearby galaxies to higher redshifts. Comparing derived parameters before and after redshifting provides a test for biases that may be introduced by degraded resolution and cosmological surface brightness dimming. This technique has been used in the past for different purposes, for instance assessing morphologies at higher redshifts \citep[e.g.,][]{petty2009, conselice2011, vanderbergh2002, lisker2006, giavalisco1996}. Recently, \cite{barden2008} used a set of $\sim100$ local  galaxies to study the cosmological redshifting effect on size and shape of galaxies at $0.1<z<1.1$. They created new images from the best-fit single \ser models of their input images and then redshifted them to show that there are no systematics on the size and morphological parameters. However, nearby galaxies have signs of different sub-structures and low surface brightness features. Generating simulated galaxies with a comparable range of properties of galaxies and adding them into the blank sky background images is a practical test. However, these mock objects are simple cases compared with real objects and could be assumed to produce  lower limits on the systematics \citep[e.g.,][]{trujillo2006a}.

It is also a common practice to measure the surface brightness profile of galaxies at high-\textit{z} using single component \ser profile fitting. Therefore, it is assumed that for a consistent comparison of sizes at low and high-\textit{z}, the profiles of nearby galaxies also should be measured with the same method. However, as mentioned earlier, galaxies often consist of multiple components (i.e., bars, bulges, compact cores, spiral arms, etc.). In the local universe, these sub-components are well-resolved and distinguishable in the photometric analysis of their structures. Therefore, their surface brightness profiles may deviate from a single component model. It has been shown that using extra components in fitting surface brightness profiles of nearby galaxies better describes the underlying stellar distributions than using canonical single \ser profile fitting \citep[e.g., for elliptical galaxies:][; also see references therein]{ferrarese1994, lauer1995, graham2003, huang2012}.  Some authors have also shown that using single component \ser profile fitting for nearby galaxies with more than one component might systematically bias sizes and morphological parameters \citep[e.g.,][]{meert2012, bernardi2012}. 

Therefore, in this paper, we first investigate the biases associated with estimating sizes of nearby galaxies using single \ser profile fitting and its effect in their comparison with galaxies at high redshifts. These effects will also be tested against various types of galaxies (e.g., classifications according to their morphology, color, star-formation rate). We will explore the possible dependence of the systematics of sizes on the galaxies classifications and test alternate (two-component and nonparametric) methods.

We also artificially redshift real images of nearby galaxies ($z\sim0$) to $z=1$ in order to investigate the uncertainties of parameter measurements. We use the resolution of \textit{HST} WFC3 instrument, since images from this instrument are now being widely used for studying galaxy structures at high redshifts \citep[etc.]{ oesch2010b, szomoru2012, patel2012, newman2012, mosleh2012, sande2012}. Moreover, for the sake of better statistics, we use a large sample of nearby galaxies ($\sim1000$ objects). 

Finally, we use our robust size measurements to study the correlation of size and stellar masses of our nearby galaxies. Galaxies can be selected or classified by means of different methods or criteria, such as morphology, color, and star-formation rate. We investigate the mass-size relation for different types of nearby galaxies at a wide range of stellar masses and test whether the selection criteria could affect the mass-size relations. These relations provide a baseline for further studies at high redshifts. We will also compare the  mass-size relations of galaxies after artificially redshifting them to $z=1$ and examine if the robustness of galaxy mass-size relations hold at high redshifts.  

We explain our sample used in this study in Section 2. The size determination methods and their systematic offsets at $z\sim0$ are explored in Section 3. The stellar mass-size relations of nearby galaxies are studied in Section 4. We describe the artificial redshifting procedure of galaxies to $z=1$ and their sizes compare with $z=0$ objects in Section 5. We discuss our results in Section 6. The cosmological parameters adopted throughout this paper are $\Omega_{m}$ = 0.3, $\Omega_{\Lambda}$ = 0.7 and $H_{0} = 70$ $km$ $s^{-1}$ $Mpc^{-1}$.\\

\begin{figure*}
\includegraphics[width=\textwidth]{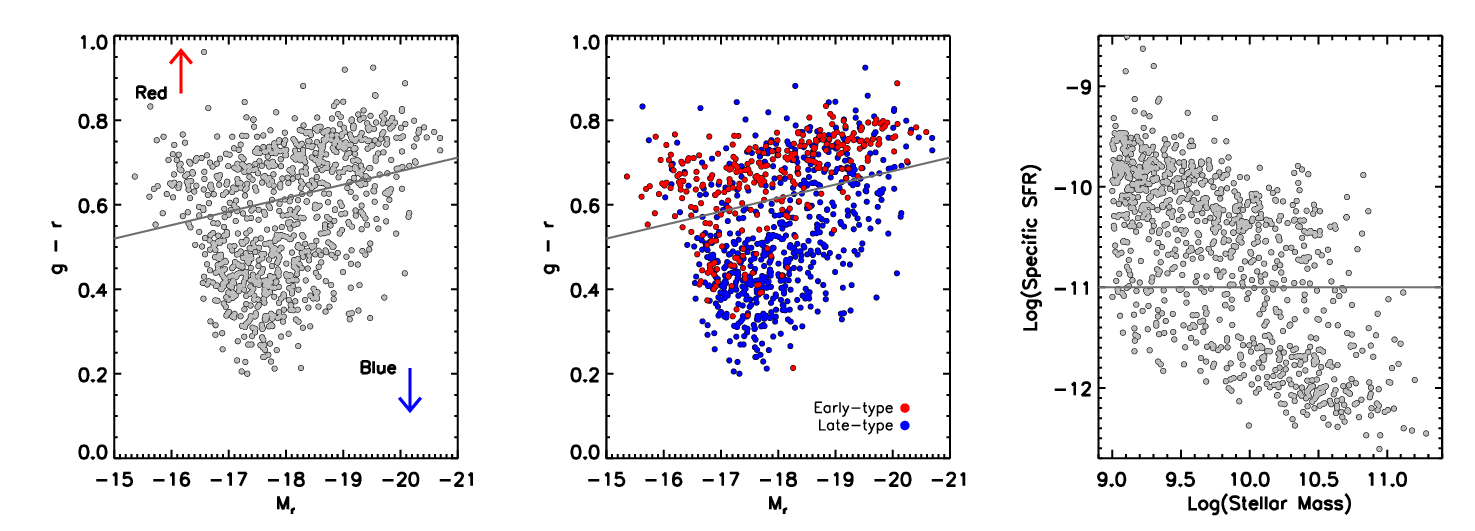}
\caption{Left: the color-magnitude diagram of our sample. The solid line shows the separating cut defining red and blue galaxies. Middle: the same as in the left panel but galaxies are color coded according to their morphological classifications, i.e., early-type galaxies are red points and late-type galaxies are blue points. The morphological classification is based on the  GZ1. Right: distribution of galaxies sSFR as a function of their stellar mass; the solid line represents the separation cut at $log(sSFR)=-11$.} 
\label{fig1}
\end{figure*}

\section{Data}

The sample of galaxies we use for this study is selected from the Max-Planck-Institute for Astrophysics (MPA)-Johns Hopkins University (JHU) Sloan Digital Sky Survey (SDSS) DR7 \citep{kauffmann2003,salim2007}, which has spectroscopic redshifts for SDSS DR7 galaxies \citep{abazajian2009} galaxies. The surface brightness limit for our sample is $\mu_{50} \leq 23$ mag arcsec$^{-2}$ with magnitude limit of $r \leq 17.77$.  We initially select galaxies to have spectroscopic redshifts within $0.01 < z < 0.02$ and stellar masses of $\log(\mstar/\msun) \geq 9$. As we intend later to artificially redshift galaxies to $z=1$, the imposed redshift limits are to avoid selecting galaxies where the SDSS point spread function (PSF) is broader than the WFC3 PSF at $z=1$ \citep[and providing sufficient sampling at high-\textit{z}; see][for more details]{barden2008} and also to avoid objects with very large apparent sizes. To reduce processing time, we further select about 1000 galaxies randomly from this sub-sample (about one third of galaxies in this mass and redshift range). We use SDSS \textit{r}-band images for measuring their sizes at this low-\textit{z}.

We classify our sample into different sub-samples based on their color, morphology, and specific star-formation rate (sSFR). The  left panel of Figure \ref{fig1} shows the distributions of all galaxies in the color-magnitude diagram. The color and absolute magnitude are based on the New York University Value-Added Galaxy Catalog \citep[NYU-VAGC;][]{Blanton2005}. Parallel to the red sequence distribution, we define the following line to separate galaxies into red and blue objects:
\begin{equation}
(g - r ) =  0.68 - 0.032(M_{r}+20)
\end{equation}
 
In order to classify galaxies based on their morphology, we used the Galaxy Zoo Catalog (GZ1) \citep{lintott2011}, which is a morphological catalog of visually classified SDSS galaxies. We classify galaxies into early-types and late-types based on the debiasing fraction of the votes for each galaxy type being dominant (see \cite{lintott2011} for more details). We note that the classification are only available for $\sim94\%$ of our sample.  The color magnitude distributions of these early-types and late-types are shown in the middle panel of Figure \ref{fig1}. Early-type galaxies are indicated as red symbols and late-type ones are shown in blue. 

Galaxies can also be selected by means of their sSFR \citep{brinchmann2004}. In the right panel of Figure \ref{fig1}, the distributions of sSFRs and stellar masses of galaxies are shown. We define $log(sSFR) = -11$ as a separating cut to split our sample into star-forming and non star-forming galaxies. In summary, we divide our galaxies by four criteria: (1) morphology based on Galaxy Zoo visual galaxy classifications, (2) color, (3) sSFR, and (4) \ser indices (based on smoothed profiles of galaxies at low-\textit{z}; see Appendix B).   

We also need to take into account the effects of sample selections on the completeness. We follow S03 to apply volume corrections to our sample (the $V_{max}$ method). We give each galaxy a weight that is proportional to the inverse of the maximum volume out to which it can be observed.  As our sample is limited to redshift ranges of $z=0.01-0.02$, all galaxies have equal weights and hence our sample is not biased by stellar mass incompleteness down to $10^9 \msun$. However, as demonstrated for example in \cite{Taylor2010}, the sample is incomplete at these redshifts due to SDSS spectroscopic selection, particularly for high-mass and/or very compact galaxies ($>10^{11} \msun$ and $<0.8$ kpc, respectively). It is worth noting that the fraction of galaxies that were not morphologically classified by Galaxy Zoo is $\sim 6\%$ on average and hence the effects are negligible. An insignificant number of galaxies (5 objects) had very large ($>6$ arcsec) offsets between the catalog position and our best-fit center, due to blending or unusually high central obscuration, and were excluded from this analysis.  \\

\begin{figure*}
\includegraphics[width=\textwidth]{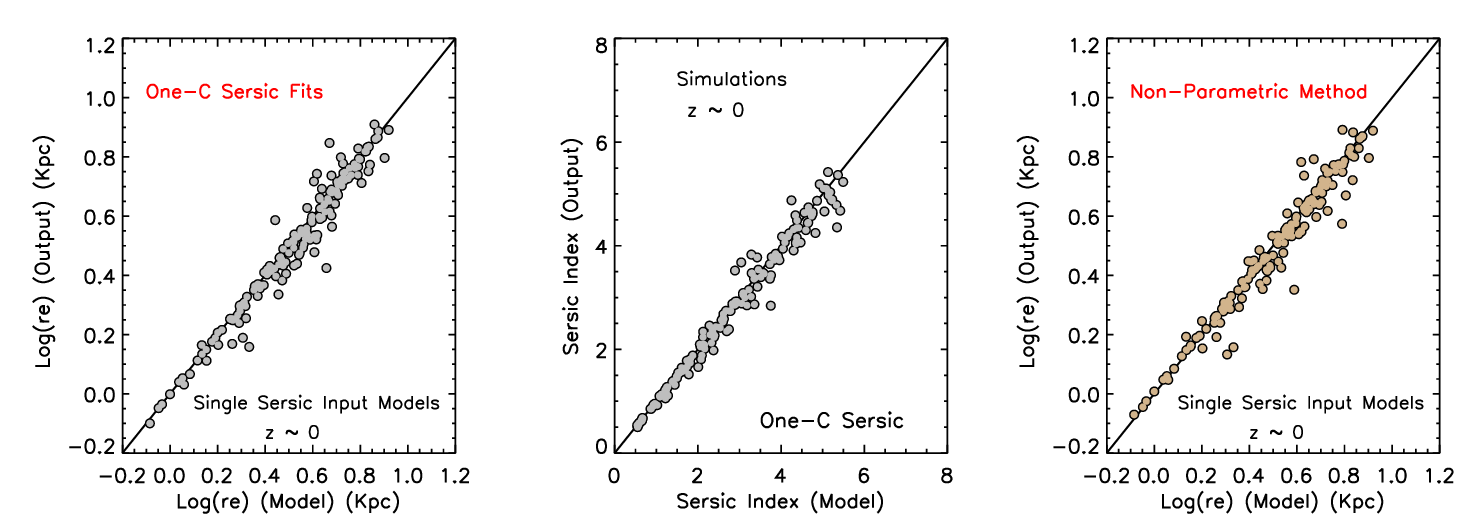}
\caption{\textit{Simulation (I)}: comparison between sizes of simulated galaxies (models with ``single'' \ser profiles)  and their recovered sizes (using single component \ser fits in the left panel and using the non-parametric method in the right panel) and \ser indices (middle panel) after adding them into empty regions of SDSS \textit{r}-band images.  As the plots show, there are no systematics in the recovery of parameters of single \ser model galaxies for both methods.} 
\label{fig2}
\end{figure*}

\begin{figure*}
\centering
\includegraphics[width=5.4in]{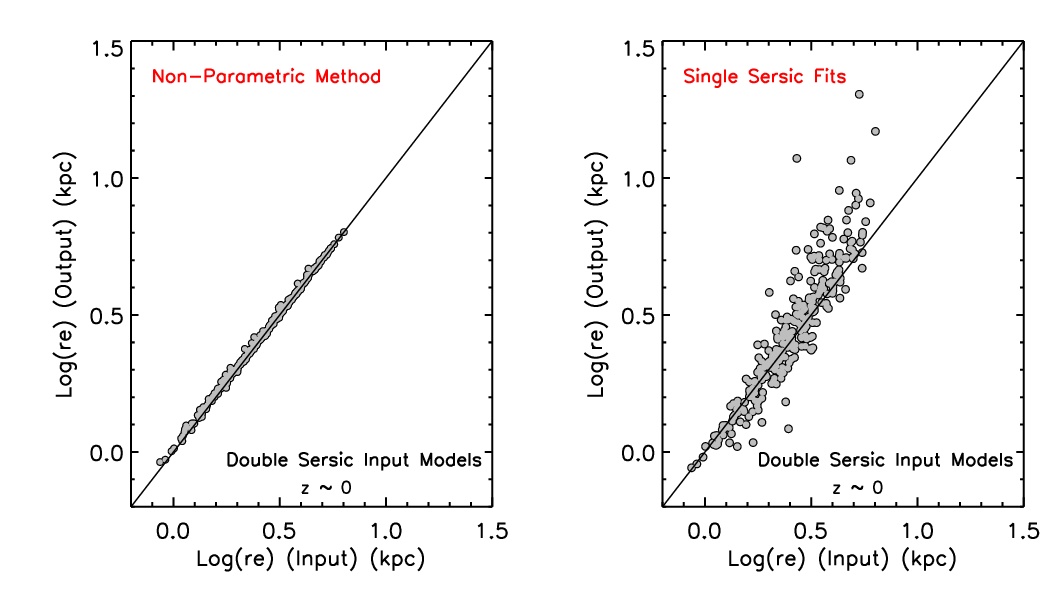}
\caption{\textit{Simulation (II)}: comparison between sizes (i.e., half-light radii) of simulated galaxies (models with ``double'' \ser profiles) and their recovered sizes using single component \ser fits (right panel) and the non-parametric method (left panel).  This shows that sizes derived from single \ser profile fitting are biased for true nearby two-component \ser profile objects. We note that this simulation does not include noise, in order isolate the biases caused by intrinsically complex structures.} 
\label{fig3}
\end{figure*}

\section{Sizes at $z=0$} 

As mentioned earlier, the well-resolved profiles of some nearby galaxies could exhibit non-\ser structures. Consequently, this raises questions about the effects of these structures on measurements of nearby galaxy sizes (i.e., half-light radii) with single-component, analytical models. In the following, we employ several methods to measure sizes of our $z=0.01-0.02$ galaxies. These methods can be separated into two main categories: ``parametric'', i.e., measuring the half-light radius of galaxies using best-fit, two-dimensional analytical models and ``non-parametric'' from their observed one-dimensional light profiles and measuring their total fluxes as described below.

\subsection{Parametric Methods}

To quantify the structural properties of galaxies with parametric methods, we use the GALFIT v3 modeling software \citep{peng2010}. GALFIT measures the shape and size of each galaxy by finding a best-fit parametric model of its two-dimensional surface brightness profile. It generates a range of profile models that are convolved with the PSF of the galaxy image and determines the best-fit model by comparing models with the galaxy light profile and minimizing the $\chi^{2}$ of the fit. GALFIT can fit one or more analytical functions such as \ser \citep{sersic1963, sersic1968}, de Vaucouleurs \citep{devaucouleurs1948}, etc., to a galaxy light profile. 

In the following, we outline the procedure for using GALFIT and measuring galaxy structural parameters from \ser models. We first created a postage stamp for each galaxy from SDSS (\textit{r-band}) imaging frames ($2048\times1448$ pixels and a pixel scale of $0.396''$). The postage stamp should be large enough to contain enough background sky pixels. We initially set our postage stamps to have widths of at least 1800 pixels. However, as our galaxies have large apparent angular sizes and they might be located at different positions on the SDSS frames, the postage stamp sizes vary a bit for each galaxy. Nevertheless, our defined box-size value creates a postage stamp for each galaxy $\gtrsim10$ times larger than the apparent galaxy sizes. These are sufficient for leaving the sky background as a free parameter during the fitting procedure. 

In order to detect and mask neighboring objects, we use SExtractor \citep{bertin1996}. For SDSS \textit{r}-band images,  we use the following  SExtractor configuration parameters for detecting sources: $DETECT\_MINAREA =10$, $DETECT\_THRESH = 1.5$, and $ANALYSIS\_THRESH =1.5$, and $DEBLEND\_MINCONT= 0.095$.  In addition, we further smoothed out the mask map created by SExtractor to reduce the plausible bias of sky background estimations from the contribution of undetected low flux regions around nearby sources. We also provide the initial parameters for GALFIT, such as half-light radius, magnitude, position angle (P.A.), and axis ratio derived from SExtractor and initially set the \ser index to a value of 2.   

The SDSS photo pipeline generates a synthesized PSF image at the central position of each galaxy using a published tool of Read Atlas Images \footnote[1]{\url{http://www.sdss.org/dr7/products/images/read\_psf.html}}. We use this code and extracted PSF images in the SDSS \textit{r}-band  for each galaxy, separately. The PSF images are required by GALFIT to convolve model images during the fitting procedure.

\subsubsection{Single Component \ser Profiles} 

Our first adopted parametric model for describing the galaxy surface brightness is the one component \ser model.  Single component \ser profiles are widely used for determining galaxy structures and properties, especially for high redshifts galaxies. The \ser function describes the surface brightness of a galaxy at radius \textit{r} as
\begin{equation}
 \Sigma (r) = \Sigma_{e} e^{-b_{n}[(r/r_{e})^{1/n} -1]}
\end{equation}

 where $r_{e}$ is the half-light radius and $\Sigma_{e}$ is the surface brightness at $r_{e}$. The shape of the galaxy profile is determined by the \ser index \textit{n}, and the value of $b_{n}$ is coupled to \textit{n} \citep[see][for more details]{graham2005}.

\subsubsection{Two-component \ser Profiles} 

Although the single \ser profile describes the surface brightness of galaxies over a large dynamic range remarkably well \citep[e.g.,][]{kormendy2009}, departures from the simple models can be used for diagnosing the formation of galaxies. Specifically, nearby elliptical galaxies tend to show either ``extra light'' or ``missing-light''  in their central regions, depending on their luminosity, and different empirical functions (e.g., ``core-\ser '' or ``Nuker'' law) have been used and suggested to parameterize these distinct components \citep{ferrarese1994, lauer2007, lauer1995, graham2003, cote2006, hopkins2009a}. However, as our sample consists of a wide ranges of luminosities and morphologies, we use double \ser profiles which allow a variety of possible inner and outer profiles for each object \citep[see][]{turner2012}. Our adopted multi-component model is described as:

\begin{equation}
 \Sigma (r) = \Sigma_{e1} e^{-b_{n1}[(r/r_{e1})^{1/n1} -1]} + \Sigma_{e2} e^{-b_{n2}[(r/r_{e2})^{1/n2} -1]}
\end{equation}

To compute effective radii, we first analytically reconstructed the sum of the deconvolved circularized surface brightness profiles of two components from the best-fit parameters and then computed their total fluxes and consequently their half-light radii.

\subsection{Non-parametric Method}

We test the results from these analytical models against an independent, non-parametric method. The non-parametric technique does not rely on previous assumptions about the structure of galaxies. It benefits from the galaxy observed curve of growth. In brief, the observed intensity profile of a typical galaxy is measured through elliptical isophotal fitting and from that, the growth curve of galaxy fluxes is determined. This provides the radius at which the flux reaches half of the total value. 

In detail, in order to measure the half-light radii of galaxies from this method, we need to integrate the fluxes of galaxies at different radii and find the radius at which the flux reaches half the value. For this purpose, we first extract the observed surface brightness profile of galaxies using the IRAF task ELLIPSE \citep{jedrzejewski1987}. This procedure measures fluxes in isophotal ellipses over the galaxy image and therefore can generate one-dimensional surface brightness profiles of  galaxies. 

The accuracy of this method depends on the precise measurements of galaxy total fluxes. Therefore, we measure the fluxes out to $\sim400$ arcsec from the galaxy centers. However, the surface brightness of galaxies is low in the outer parts and hence it is very difficult to define the exact edges of galaxies. Therefore, for measuring the fluxes in the outer parts, we extrapolate the total light of  galaxies beyond their $petroR90$ radius (i.e., a radius containing $90\%$ of the \textit{Petrosian} flux derived from SDSS DR7). This is done by fitting one-dimensional \ser profiles to these outer regions. By integrating the light profiles estimated from our best-fit models to infinity, the total fluxes in the outer regions are estimated. Moreover, in this way, we also estimate the sky background for each galaxy as the sky value is left as a free parameter during the fitting procedure.  Then, for each galaxy, we integrate fluxes at different radii up to radius smaller than $petroR90$ from the fluxes measured by ellipse fitting and add them to the fluxes estimated in the outer region. This sum represents the total galaxy flux and we use this to measure the radius within which half of the flux is contained (here referred to as the ``non-parametric'' size). We note that for approximately $7\%$ of the galaxies, the one-dimensional fits to the outer parts using $petroR90$  did not converge. For this small subset of objects, we instead perform a \ser extrapolation outside $petroR50$, which is the radius containing $50\%$ of the flux within the \textit{Petrosian} flux. In order to check if the results depend on the choice of radius for the rest of the sample, we repeated the procedure by fitting the outer parts of galaxies starting at smaller radii of, i.e., $petroR50$. The results were perfectly consistent for all galaxies, so we conclude that the choice of extrapolation radius does not affect our non-parametric sizes. We note that we fixed the ellipticity (\textit{E}) and the \textit{P.A.} of the ellipse isophotes to the values obtained from the best-fit of single \ser parametric method. 

The sizes derived from the non-parametric method also need to be corrected for PSF broadening and therefore we use the relation according: $R = \sqrt{(r_{1/2})^2-(r_{PSF})^2}$, where $r_{1/2}$ and $r_{PSF}$ are the derived non-parametric half-light and PSF size,  respectively. This correction is a crude approximation assuming Gaussian galaxy profiles; although its effect is negligible for the bulk of our sample, there could be potential systematics in the sizes of extreme galaxies with high concentrations (high \ser indices) and very small sizes ($\lesssim 1$ kpc).  It is also worth noting that all sizes derived in this paper are circularized, using $\sqrt{ab}$, in which $a$ is the semi-major axis and $b/a$ is the axis ratio. This removes the effects of ellipticity \citep[e.g.,][]{trujillo2006a, franx2008, williams2009}. \\

\begin{figure*}
\centering
\includegraphics[width=6.4in]{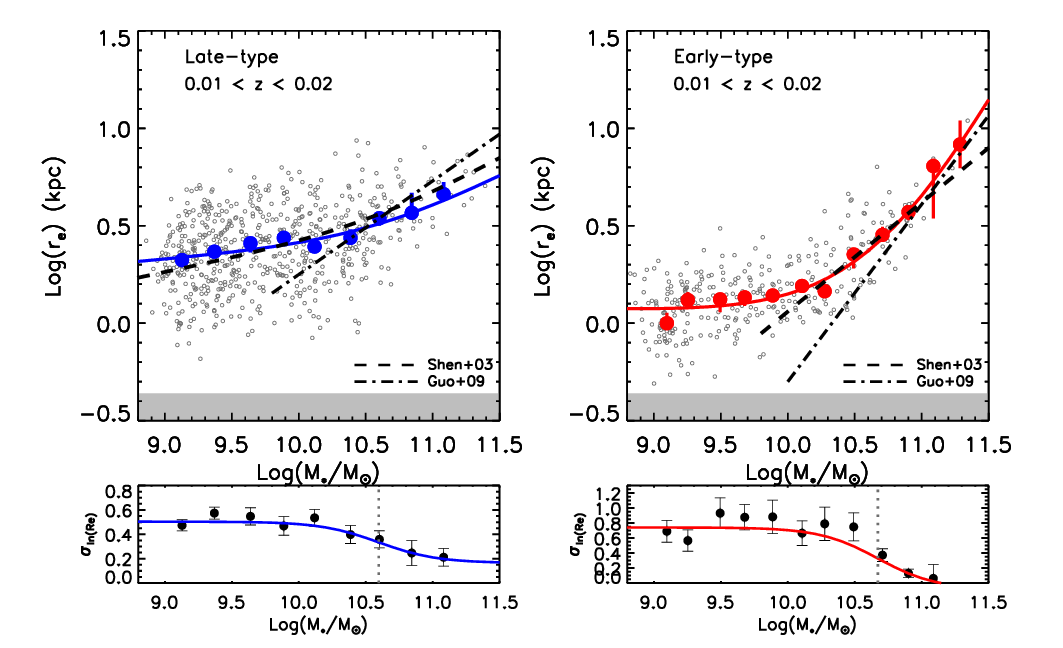}
\caption{Top row: the stellar mass-size relation of late-type galaxies (left panel), and early-type galaxies (right panel). The individual galaxies are shown as small open gray circles and the blue and red filled circles are the median of the sizes in stellar mass bins.  The solid blue and red lines are the best fits to the data. The mass-size relation from studies of SO3 and \cite{guo2009} are also illustrated by dashed and dot-dashed lines, respectively. The best-fit relations are consistent with SO3, however the relation flattens below $M\lesssim \sim 4\times10^{10} \msun$ for early-type galaxies. The shaded gray regions show the physical sizes of PSFs in the SDSS r-band images. Bottom row: the size dispersion as a function of stellar mass and their best-fits. The characteristic stellar masses, where the dispersions change significantly, are shown as vertical dotted lines.} 

\label{fig4}
\end{figure*}

\subsection{Simulations \textit{(I) and (II)}}

We perform simulations for testing our methods and procedures, as follows. The first simulation is designed to test the reliability of the single component \ser method and the non-parametric method for galaxies at $z=0$. For that, we first generated single \ser mock galaxies with random properties (magnitude, re, b/a) with a similar range of values as real galaxies ($11.5 < $mag$ < 17.7$,  $0.4$ kpc $ < re < 10.9 $ kpc ,  $0.2 < $ b/a $ < 1$). We then add them into the empty regions of the $r$-band SDSS images and perform our fits, using single \ser profiles and the non-parametric method. The results are shown in Figure \ref{fig2}. In the left panel, the comparison between input and output sizes is shown using the parametric method; the \ser indices are compared in the middle panel. The output sizes derived by using the non-parametric method are also compared with their original sizes in the right panel of Figure \ref{fig2}. The sizes of galaxies can be recovered without any systematics with median differences of less than $2\%$ for both methods. There are also no systematic errors in the recovery of the \ser indices. This shows that  our procedure is robust for the recovery of mock galaxy properties, assuming single \ser profiles and using SDSS images.

As shown in Appendix A (Figure \ref{fig10}), the sizes of galaxies derived using single \ser profile fitting can be biased, especially for massive early-type galaxies. This may be caused by the existence of additional component(s) or non-\ser light profiles.  We have also shown that sizes derived with double \ser components, are smaller that the sizes from the one-component models. We test an idealized case using simulated two-component objects. For that, we first created a sample of 300 two-component \ser galaxies such that each model galaxy has a central component with median half-light radius of $\sim1$ kpc and an outer component with a median size of $\sim3$ kpc. We also assumed that the central components have larger median \ser indices than the outer-part components. For all galaxies, the central components are $\sim 0.6$ magnitude fainter than the outer components. These numbers are derived from the average results of the two-component fits to our real galaxies at $z=0.01-0.02$.  To ensure that we are testing only the effects of multi-component galaxies, only the sky background levels are added to the images of these model galaxies without any additional noise or neighboring objects. We then measure the sizes of these two-component model objects using single \ser profile fitting and the non-parametric method. The results are shown in Figure \ref{fig3}. As seen in the left panel, the sizes are recovered robustly with the non-parametric method. However, as shown in the right panel of Figure \ref{fig3}, the sizes from single \ser fitting are biased (larger) compared with their input half-light radii, especially for large objects. This simplified test shows that sizes from single \ser profile fitting can be biased for true two-component galaxies.  \cite{meert2012} use different assumptions for simulated SDSS galaxies and show the existence of a bias in the recovered parameters when fitting a single \ser profile to real two-component systems.  Although our sample is comprised of quite nearby objects ($\sim 45-85$ Mpc), \cite{bernardi2012} show the same effect for the main SDSS sample at $z\sim0.1$. Hence, using single \ser sizes for local galaxies can introduce systematics in size analyses. 

Nevertheless, fitting correct models to nearby galaxies is complicated. Different authors use different models to fit multi-component galaxies, e.g., traditional deVaucouleurs plus an exponential disk,  \ser + exponential \citep{meert2012}, double \ser or even using multiple (3-4) \ser profiles \citep{huang2012}. It is also the case that not all of the galaxies (at wide ranges of stellar masses) need to be measured by multi-component models ($\sim77\%$ are robustly fit with two-component models in this work). Therefore, for the rest of this study, we use our non-parametric sizes for these $z\sim0$ galaxies. Our simulations (\textit{I} and \textit{II}) demonstrate the robustness of our non-parametric method. In addition, due to the  large angular sizes of our galaxies, the effects of the PSF on sizes from this method are negligible.

It is worth noting that fluxes used for estimating the stellar masses of SDSS galaxies are model dependent and hence these fluxes can be different from fluxes measured using the non-parametric method for each individual galaxy. Therefore, it is essential to correct the stellar masses according to the new flux measurements. We rescale the stellar mass of each galaxy by measuring the ratio between its non-parametric flux and the flux used for estimating its stellar mass from the MPA catalog. Comparing the rescaled stellar masses with the ones from the MPA catalog shows that there are no systematic differences for stellar masses $\log(\mstar/\msun) < 10.7$, increasing to at most $+0.1$ dex for $\log(\mstar/\msun) > 11$. This mass rescaling, while formally correct, therefore does not substantively affect our results.\\

\begin{figure*}
\centering
\includegraphics[width=6.4in]{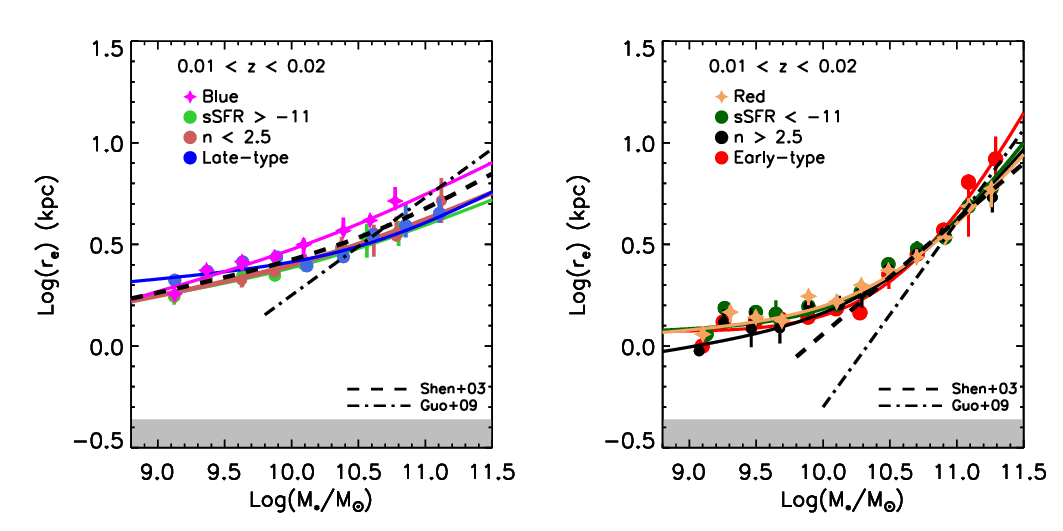}
\caption{Stellar mass-size relation of galaxies classified by means of different criteria. In the left panel, blue, late-type (visually classified), star-forming galaxies, and low-\ser index ($n <2.5$) systems are compared and in the right panel red, early-type (visually classified), non star-forming, and $n>2.5$ systems are compared. The stellar mass-size relation from studies of SO3 and \cite{guo2009} are also illustrated by dashed and dot-dashed lines, respectively. The points are the median size dispersions as a function of stellar mass and the lines represent the best-fits to these points. As this plot shows, the relations based on different methods of classification of galaxies are largely consistent, although blue galaxies lie above the other relations.}
\label{fig5}
\end{figure*}

\begin{deluxetable*}{ccccccc}
\tablecolumns{7}
\centering
\tablewidth{\textwidth}
\tabletypesize \footnotesize
\tablecaption{The Fitting Results of the Parameters in the Size-Mass Relations.}
\tablehead{
\colhead{Sample} &
\colhead{$\alpha$} &
\colhead{$\beta$} & 
\colhead{$\log(\gamma)$}&
\colhead{$M_{0}$}&
\colhead{$\sigma_{1}$}&
\colhead{$\sigma_{2}$}
}
\startdata
   Early-type               & $-0.020\pm0.077$  &   $1.258\pm0.210$	  & $0.247\pm0.734$  &   $10.673\pm0.202$ & $0.741\pm0.078$ & $-0.085\pm0.247$\\
   Red                         & $0.042\pm0.051$    &   $0.802\pm0.126$    & $-0.314\pm0.479$ &  $10.537\pm0.131$  & $0.758\pm0.092$ & $0.130\pm0.077$\\
  $ log(SSFR) < -11$   &  $0.014\pm0.069$  &   $0.912\pm0.168$   & $-0.058\pm0.652$   & $ 10.555\pm0.107$ & $0.869\pm0.114$ & $0.130\pm0.081$\\
   $n$ $ > 2.5$           &  $0.094\pm0.096$  &   $0.829\pm0.214$   & $-0.864\pm0.905$  &  $10.531\pm0.166$  &  $0.751\pm0.155$ & $0.148\pm0.094$\\
   Late-type                 &  $0.058\pm0.059$  &   $0.357\pm0.181$    & $-0.197\pm0.548$   &  $10.597\pm0.233$    &  $0.503\pm0.041$  & $0.164\pm0.105$\\
   Blue                          &  $0.185\pm0.081$  &   $0.329\pm0.164$    & $-1.406\pm0.750$   &  $10.325\pm0.299$   &  $0.574\pm0.059$ &  $0.056\pm0.122$\\   
  $log(SSFR) > -11$	     &  $0.109\pm0.090$  &    $0.263\pm0.196$   & $-0.743\pm0.831$     &  $10.204\pm0.214$   &  $0.668\pm0.046$  & $0.234\pm0.100$\\
  $n$ $< 2.5$              &   $0.124\pm0.081$  &   $0.278\pm0.161$    &  $-0.874\pm0.756$   &  $10.227\pm0.230$  &  $0.671\pm0.054$  & $0.249\pm0.091$\\

\enddata
\tablecomments{The best-fit parameters for the stellar mass-size relation for different types of galaxies at $z\sim0$ (Equations (4) and (5)). }
\end{deluxetable*}


\section{Stellar Mass-Size Relation At $z=0$}

The stellar mass-size relation for SDSS galaxies has been studied by S03. They investigated this relation for objects that are defined as early- and late-types according to their \ser and concentration indices and their relations have been widely used in literature. However, it is argued that the half-light sizes used in S03, which are from the NYU-VAGC catalog and based on one-dimensional single \ser fitting,  could have been underestimated \citep[e.g.,][]{guo2009, simard2011}. We have also shown that using single \ser fitting could bias the sizes of galaxies with high stellar masses. As the mass-size relation could depend on the fitting model employed (specifically at the high-mass ends), our independent non-parametric method for $z=0$ galaxies, should remove uncertainties due to model assumptions.  Our sample consists of galaxies over a wide range of stellar masses ($\gtrsim 10^{9} \msun$) and is suitable for investigating this relation. 

We first study the mass-size relation of our sub-samples based on morphological Galaxy Zoo classifications. The distribution of sizes versus stellar masses of late-type and early-type galaxies is illustrated in the top row of Figure \ref{fig4}. The median sizes in small bins of stellar masses for each sample are measured (blue and red circles) and it can be seen that sizes of both late-types and early-types show little correlation with masses up to $\sim3-4 \times10^{10} \msun$; however, the relation steepens beyond this stellar mass and is stronger for  early-type galaxies. For both types of galaxies, the relations seem to begin above specific stellar masses. 

To further quantify the correlations, we use the functional form employed for late-type galaxies in S03 (Equation (18)) for both our late-type and early-type samples:

\begin{equation}
R_{kpc} = \gamma (\mstar/\msun)^{\alpha} (1+\mstar/M_{0})^{\beta-\alpha}
\end{equation}

where, $\alpha$, $\beta$, $\gamma$, and $M_{0}$ are free fitting parameters. This basically allows the relation to have two different slopes depending on the stellar mass range. $\alpha$ and $\beta$ represent the slopes of the relation, and the characteristic mass, $M_{0}$, determines the stellar mass at which the slope of the relation changes. However, this relation is not very sensitive to the characteristic mass, $M_{0}$, therefore, this can be defined from the size dispersion relation as follows (Equation (19) in S03):

\begin{equation}
\sigma_{ln R} = \sigma_{2} + \frac{(\sigma_{1}-\sigma_{2})}{1+(\mstar/M_{0})^2} 
\end{equation}

where $\sigma_{1}$ and  $\sigma_{2}$ are also free fitting parameters (representing the size dispersions at low and high masses) and  $M_{0}$ is the characteristic stellar mass at which $\sigma_{ln R}$ significantly changes. Size dispersions as a function of stellar mass for late-type and early-type galaxies are shown in the bottom panels of Figure \ref{fig4} (left and right panels, respectively). The best fits to the data points are shown as solid blue and red lines and the best-fit parameters are presented in Table 1. 

For late-type galaxies, the median size dispersions decrease at stellar masses greater than $\sim4\times10^{10} \msun$, consistent with S03.  The mass-size relation for these galaxies is also consistent with S03 (dashed line).  The size dispersions for early-types also behave similarly and decrease for massive galaxies above a characteristic mass around $4\times10^{10} \msun$. However, due to low number of objects in these high-mass bins, it is not clear how significant this effect is.

The median sizes of early-type galaxies in a stellar mass range of $\log(\mstar/\msun) \sim 10-11$ are consistent with the S03 relation. However, at lower stellar masses ($\lesssim 2\times10^{10} \msun$), the sizes are almost constant. Therefore, in this mass range, there is little correlation between stellar mass and size. However, we caution that the flattening in this relation below $\log(\mstar/\msun) \sim 9.5$ may be in part due to systematic effects, since a significant fraction of quiescent galaxies in this mass regime have sizes comparable with the PSF. For late-type galaxies, the relation runs parallel at these masses but with larger sizes. The mass-size relations for galaxies with higher stellar masses (i.e., $\gtrsim 2\times10^{10}\msun$) are steep for both late- and early-types. However, each sample exhibits different slopes and early-types have a steeper mass-size relation (see Table 1). 

We also present the mass-size relations of galaxies based on different sample definitions such as color, \ser indices, and sSFR in Figure \ref{fig5} in order to test the effects of these selections on the mass-size relation and defining baselines for future studies based on different sample classifications. Interestingly, the mass-size relations based on these classifications are  consistent with the analogous relations in Figure \ref{fig4}. In the left panel of Figure \ref{fig5},  late-type galaxies are compared with star-forming, blue, and low \ser index galaxies. They are almost consistent, although the blue galaxies have larger sizes at stellar masses $\gtrsim 10^{10}\msun$ compared with the others. This could be caused by excluding edge-on galaxies using the color criterion.  We should note that the \ser indices are measured from the degraded and smoothed SDSS images of galaxies (see Appendix B, Figure \ref{fig15}), hence removing biases from sub-structure. Nevertheless, it is interesting that the mass-size relation obtained for objects with \ser indices $ n<2.5$ are consistent with S03. The best fits to the mass-size relation are summarized in Table 1. 

The right panel of Figure \ref{fig5} illustrates the comparison between the mass-size relation of early-type galaxies and those with red colors, low sSFRs, and high \ser indices. The relations are also consistent with each other. For all samples, the relations are curved with a weak relation for galaxies below  $4\times10^{10} \msun$.  The slopes of the mass-size relations at the high mass ends ($\beta$) for these red/quiescent/$n>2.5$ galaxies are on average around $\sim 0.85$, close to the slope of early-type central galaxies in \cite{guo2009}. However, this slope is slightly larger for early-type galaxies.  

In general, we show that the stellar mass-size relations for both late- and early-type galaxies are curved with a steeper slope at higher stellar masses. The size dispersions below the characteristic masses are high but decrease above $M_{0}$. This is the case for all of our studied samples.  The stellar mass-size relations based on different definitions, such as color, sSFR, and morphology, are consistent with the scaling relations of late- and early-type galaxies. We note that more restrictive sample definitions, e.g., choosing higher and lower sSFR thresholds for star-forming and quiescent galaxies, do not qualitatively change the results.\\

\begin{figure*}
\centering
\includegraphics[height=3. in]{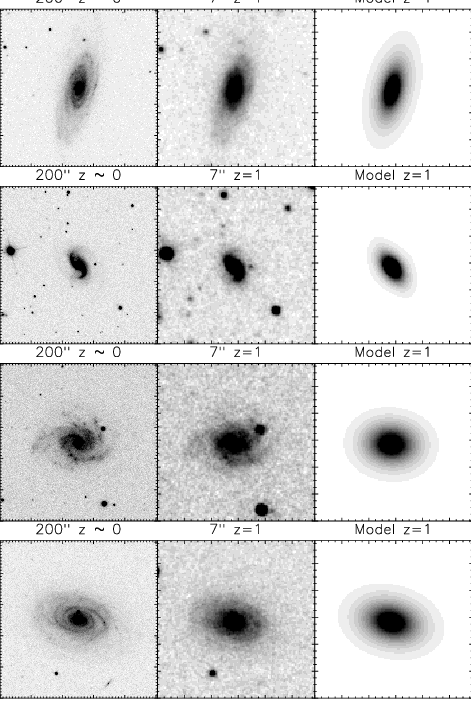}
\hspace{10. mm}
\includegraphics[height=3. in]{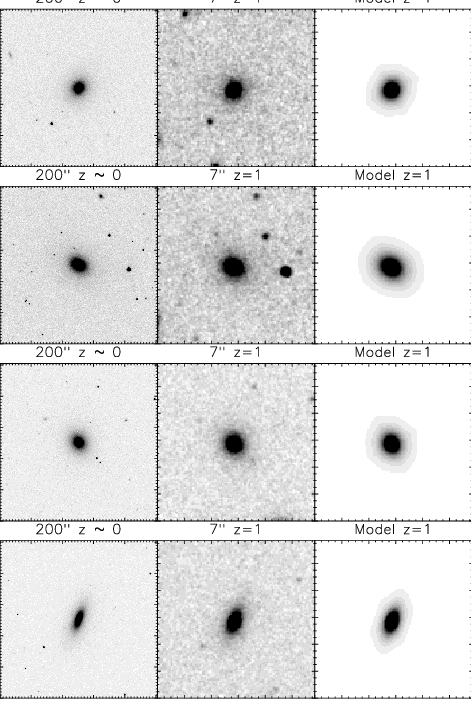}
\caption{Example images of four spiral galaxies (left set of panels) and four elliptical galaxies (right set of panels). In each set, the left panels show the SDSS (\textit{r-band}) postage stamp images ($200''\times200''$) of galaxies  at $0.01 < z < 0.015$. In the middle columns, we show their artificially redshifted (to $z=1$) postage stamp ($7''\times7''$) images after adding the WFC3-\textit{$J_{125}$} images. The right columns show the best-fit single \ser models of these redshifted galaxies.}
\label{fig6}
\end{figure*}

\begin{figure}
\centering
\includegraphics[width=3.4in]{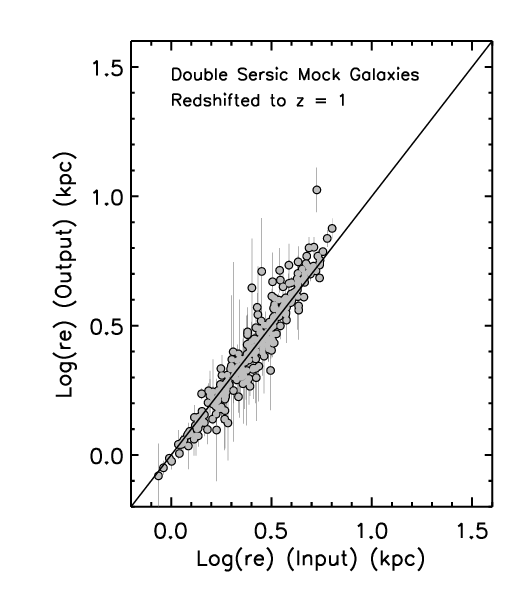}
\caption{Sizes of simulated two-component $z\sim0$ galaxies that have been ``redshifted'' to $z=1$ and re-measured with single-component \ser profile fitting.  The input and output sizes are consistent, indicating that sizes of multi-component galaxies can be reliably derived with single-component \ser models at higher redshifts.}

\label{fig7}
\end{figure}

\section{Redshifting Galaxies to $z=1$}

In order to check whether cosmological effects and observational uncertainties could affect size (and structural parameter) measurements of galaxies at high redshifts, we perform redshifting simulations of the low-\textit{z} objects. We use our sample of galaxies from SDSS at $z\sim0$ to create artificially redshifted samples of galaxies resembling the same galaxies at $z=1$ in the \textit{HST} WFC3 images. Our redshifting procedure is similar to the method described by \cite{barden2008} (FERENGI code) and we briefly describe it below. However, in order to take into account the effects of bandpass shifting, we only use SDSS \textit{r}-band images as input and use WFC3- $J_{125}$ images from the CANDELS DEEP DATA \citep{bouwens2012} as output images instead of using the \textit{k}-correction method described in \cite{barden2008}. WFC3 is the new near-IR instrument on board  \textit{HST} and covers rest-frame optical wavelengths at $z\sim1-3$. Hence, it is suitable for this purpose. 

\subsection{Method}
The first step in the redshifting procedure is to re-bin the low-$z$ images with pixel scale  $p_{i}$ and redshift $z_{i}$ to output images at redshift $z_{o}$ ($=1$ in this work) and pixel scale $p_{o}$ by a factor of $\beta$ as 

\begin{equation}
\beta = (\frac{D_{i}}{D_{o}})( \frac{p_{i}}{p_{o}})   
\end{equation}

where $D$ is the angular diameter distance, expressed as $D = \frac{d}{(1+z)^{2}}$, and $d$ is the luminosity distance.  

The next step is to apply cosmological surface brightness dimming at the rate of $(1+z)^{-4}$ in each re-binned pixel. By considering the fact that the absolute magnitude of galaxies must be conserved, the total fluxes $f$ of the input and output images must scale as: 

\begin{equation}
(\frac{f_{o}}{f_{i}}) = (\frac{d_{i}}{d_{o}})^2
\end{equation}

We note that it has been shown by several studies \citep[e.g.,][]{barden2005, labbe2003} that the intrinsic surface brightness of galaxies increases with redshift. Therefore, during our procedure of artificially redshifting our galaxies, we incorporate the surface brightness evolution, making the galaxies one magnitude brighter at $z=1$, following $M_{evo} = xz + M$  and setting $x=-1$ \citep{barden2008}. 

It is important to replicate the same resolution of real data at high-\textit{z}. Therefore, the next step is to correct the images to the appropriate PSF. This can be done by finding suitable kernels for convolving low-\textit{z} images to reach the same PSF properties/shape at high-\textit{z}. To do this, for each galaxy we require two PSFs, i.e., its low-\textit{z} and high-\textit{z} PSFs.  We use low-\textit{z} PSFs from SDSS (the ones we used for measuring sizes at $z\sim0$) and the median-stacked PSF, which is made from non-saturated stars in the $J_{125}$ WFC3 images, for the high-\textit{z} PSF,. Then, by transformation of PSFs into Fourier space, finding their ratio, and transforming the results back into spatial domain, we can find the convolution kernels required to reach the WFC3's $J_{125}$-band PSF.  Note that we calculate separately a transformation function for each galaxy as the kernel depends on the input and output redshifts.

After transforming images to the high-\textit{z} resolution and pixel scale, the last step is to add background noise to the images. For this, we put galaxy images into random empty regions of the $J_{125}$-band CANDELS DEEP images and then measure their structural parameters, as described below.  We note that, in order to check the effects of sky variations on galaxy property measurements, we repeated this step by inserting each redshifted galaxy into multiple empty regions. The final measured size/parameter for each object is the median of seven realizations. 

The procedure to measure the structural properties of artificially redshifted galaxies (i.e., size and \ser index) is similar to that used in \cite{mosleh2012}. In brief, we used GALFIT  to find the best-fit single \ser model for each galaxy. Neighboring objects are detected by running SExtractor and masked during profile fitting. Initial parameter guesses, such as magnitude, half-light radius, and axis ratio, are provided from the SExtractor output. We used the median-stacked PSF from stars in the field. In Figure \ref{fig6}, we show the SDSS postage stamp images of the galaxies (late-types in the left set of panels and early-types in the right set of panels) at low-\textit{z} (the left columns) and after redshifting to $z=1$ (middle columns).  The best-fit single \ser models of these artificially redshifted galaxies at high-\textit{z} are shown in the right columns. 

We perform two sets of simulations to test the size measurement accuracy in the $J_{125}$ WFC3 images and check the procedure for artificially redshifting the galaxies. These tests are described in Appendix B. We show that our redshifting method and size measurements at high-\textit{z} are robust and can recover sizes and structural parameters of model galaxies without any systematics. 

However, as discussed  earlier, using single \ser profile fitting for more complex galaxies in the nearby universe potentially biases size estimates. This fact raises concerns about the sizes of galaxies at high redshifts derived from single \ser fitting. Therefore, it is also worth checking whether single \ser profile fitting biases sizes of two-component objects at high redshifts. For that, we use the same simulated two-component model galaxies in Section 3.3 (simulation (II), Figure \ref{fig3}) and redshift them to $z=1$. We measured their sizes after redshifting with single \ser models.  The results are shown in Figure \ref{fig7}. This shows that single \ser profile fits of two-component galaxies at $z=1$ provide reliable sizes, likely due to the smaller structures being washed out at high redshift. Therefore, traditional single \ser surface brightness fitting robustly recovers sizes of our redshifted galaxies.\\

\begin{figure*}
\centering
\includegraphics[height=3.5 in]{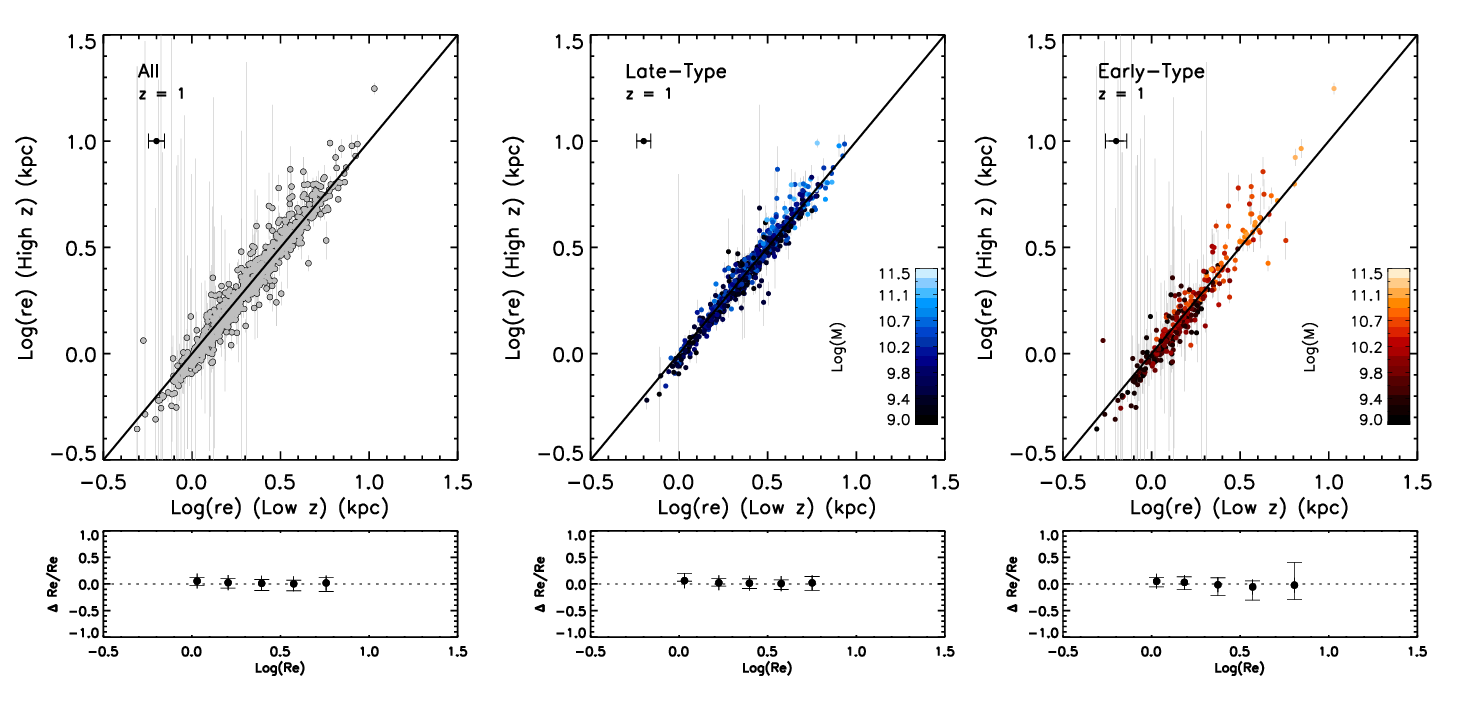}
\caption{Comparison between the sizes of galaxies at low redshift and their sizes measured after artificially redshifting all galaxies to $ z=1$ (left panel), late-type galaxies to $z=1$ (middle panel), and early-type galaxies to $z=1$ (right panel). The errors are the standard deviation of their sizes measured at different positions (different realizations). The sizes of galaxies are recovered after redshifting without any systematics. Note that due to small-number statistics, the average values of $\Delta (re)/re$ for galaxies with $re<1$ kpc in the lower panels are not illustrated.}
\label{fig8}
\end{figure*}

\begin{figure*}
\centering
\includegraphics[width=6.4in]{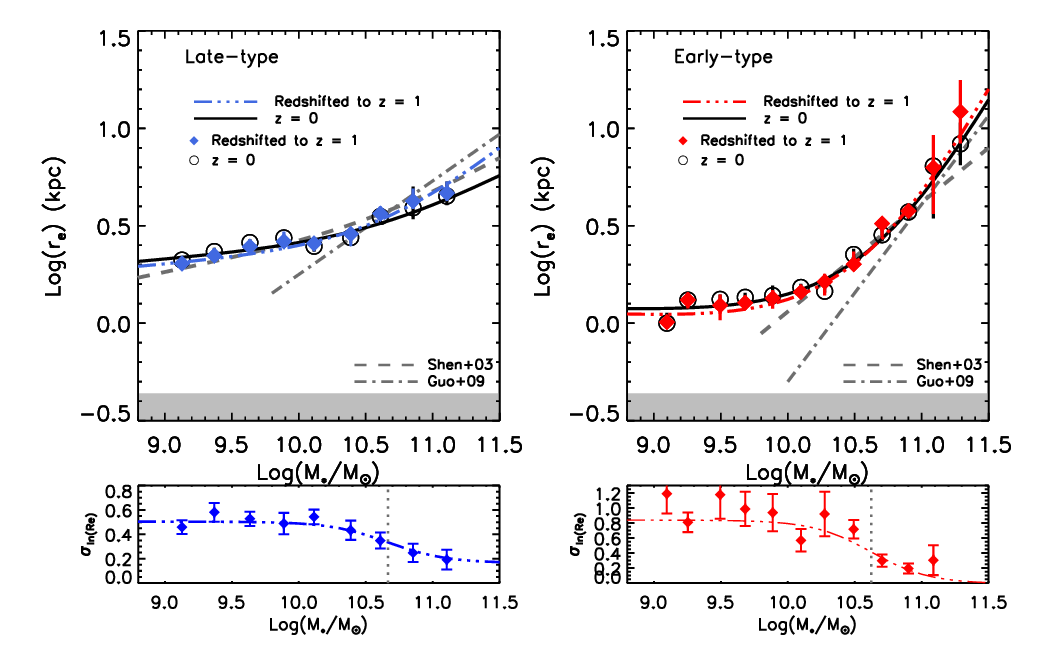}
\caption{Stellar mass-size relations and size dispersions of late-type (left panels) and early-type (right panels) galaxies are compared before and after artificially redshifting to $z=1$. The blue and red points are the median sizes of galaxies in mass bins and the dashed-three dotted lines are their best fits. The mass-size relations are consistent with the relations at $z=0$ (open circles and solid lines). This further demonstrates that size-mass relations of galaxies are reliable at $z=1$ using single \ser profile fitting.}
\label{fig9}
\end{figure*}

\subsection{Comparing with Sizes at $z=0$}

In previous sections, we described how the sizes of our sample are measured reliably at both $z\sim0$ and $z=1$. In this section, we compare sizes of galaxies before and after redshifting to $z=1$. The comparison between low-\textit{z} and high-\textit{z} sizes for all galaxies are shown in the upper-left panel of Figure \ref{fig8} and their median relative differences in small bins of sizes are shown in the bottom-left panel. As can be seen, sizes before and after redshifting agree well and there are no systematics.

There are also no biases if we split the sample into blue and red galaxies. Although the random scatter increases with size for red objects, there are no systematics and the sizes of these galaxies can be reliably recovered at high redshift, on average. As discussed in Section 5.1, we perform different realizations by inserting the galaxies into different random blank sky regions and re-measuring their properties to check the effects of the sky background on the properties of galaxies at $z=1$,. The galaxy parameters at $z=1$ are the median values of these repeated measurements and the error bars illustrate the $1\sigma$ scatter. Galaxies are also color coded according to their stellar masses. 

The results are the same using different galaxy classifications. For instance, in Figure \ref{fig8}, the size comparison is shown for late-type (middle panel) and early-type galaxies (right panel). Although the scatter increases for large and massive early-type galaxies, there are no any systematic differences in their sizes. 

It is worth noting that for sizes at $z=0$, we used the non-parametric method while we used single  \ser profile fitting for galaxies at $z=1$. Using single \ser profile fitting at $z\sim0$ results in systematics when comparing sizes before and after redshifting \citep[e.g.,][]{weinzirl2011}. This is also the case for comparing \ser indices, which tend to be overestimated at $z\sim0$ using single \ser profile fitting.

The fact that sizes of multi-component galaxies at $z=1$ can be recovered robustly using single \ser fitting can be explained by the resolution limit of images at high redshifts. The differences are mostly noticeable for massive, early-type galaxies. The bright centers of elliptical galaxies have typical sizes  $\lesssim 1$ kpc \citep[e.g.,][]{huang2012, hopkins2009a, hopkins2009b}, which is about the typical size of the PSF FWHM of the WFC3 images ($\sim1.2$kpc) at $z=1$. As a result,  the inner components are smeared out and the galaxy profiles are dominated by the outer components. Therefore, using single \ser fitting at this redshift and resolution robustly recovers the true parameters (see Appendix B for additional tests that illustrate how degrading the resolution affects the measured structural parameters of local galaxies). 

We have also checked whether the results after redshifting are sensitive to the S/N of the images. This has been tested by changing the S/N, either by adding noise to the SDSS \textit{r}-band images before redshifting them or by arbitrarily increasing the S/N of redshifted objects. These tests did not show any systematic changes in the sizes of redshifted objects. Therefore, in general, the sizes of galaxies at high redshift can be measured robustly using canonical single \ser profile fitting as long as the physical resolution is not better than $\sim 1$ kpc. \\

\subsection{Stellar Mass-Size Relation after Redshifting}

It is now be interesting to examine what the stellar mass-size relations look like after redshifting to $z=1$. In Figure \ref{fig9}, a comparison of the mass-size relations before and after redshifting to $z=1$ is illustrated. The relations for late-type galaxies are shown in the left panel, where the solid blue diamonds are the median sizes after redshifting in small mass bins and the dashed-three-dotted line is the best-fit to the data. The open circles are the median sizes at $z=0$ along with the solid black line as a best-fit  (same as in Figure \ref{fig4}). The size dispersions after redshifting are also shown in the bottom left panel. The relations for early-type galaxies are shown in the right panel.  As can be seen, the mass-size relations are consistent with their $z=0$ values after redshifting. The poor constraints at the high stellar mass end are due to small-number statistics; however, the results are consistent within the uncertainties. Figure \ref{fig9} shows that the stellar mass-size relations based on the single \ser profile fitting at high redshifts are robust. Using different definitions for separating galaxies would result in the same results after redshifting. In Figure \ref{fig9}, we only present the relations for the morphologically selected sample.\\

\section{Discussion}

In this paper, we use a sample of about 1000 galaxies at $0.01<z<0.02$ from SDSS DR7 to study their sizes and stellar mass-size relations. We first investigate the robustness of the size measurement methods for these nearby galaxies, using two main procedures for determining sizes and structures: parametric methods (single- and double-component \ser profile fittings) and a non-parametric method. In agreement with recent works \citep[e.g.,][]{allen2006, bernardi2012, meert2012, huang2012}, the majority of galaxies in the nearby Universe are well fit with two component profiles. Comparing the sizes from the non-parametric method (and from double \ser fits) with those from single \ser fits shows the systematic overestimation of sizes from the single \ser method. In particular, sizes and \ser indices of early-type galaxies at the high stellar mass end tend to be overestimated using the single \ser fitting approach. Non-\ser profiles or substructures in nearby galaxies may be the cause of this bias. We tested this by simulating two-component model galaxies and measuring their sizes using single \ser fitting and reached the same conclusion. Using single \ser profile fitting also overestimates the \ser indices of these galaxies. Therefore, we caution that relying on single \ser fits can introduce biases for nearby, well-resolved galaxies.

Stellar mass-size relations of $z=0$ galaxies from surveys like SDSS are often used as baselines for quantifying the evolution of higher redshift galaxy sizes. Using the non-parametric method and classifying our sample thorough a number of frequently employed criteria (size, color, morphology, and \ser index), we have explored the stellar mass-size relation of galaxies in the nearby universe down to a stellar mass of $10^{9} \msun$. We show that the slope of the relation varies with mass for both late-type and early-type galaxies. The relations flatten for galaxies below about $3-4\times10^{10} \msun$ \cite[see also Figure 11 in][]{turner2012}. Moreover, at these low stellar masses, the relations for both late-types and early-types run parallel but with smaller sizes for early-type objects. However, because the PSF is an increasingly significant fraction of the galaxy size at  $re < 1$ kpc, the sizes of such compact galaxies may be systematically biased. The mass-size relation for early-type galaxies below $\log(\mstar/\msun) \sim 9.5$, where significant numbers of early-types are smaller than this, should thus be considered highly uncertain. Above a characteristic stellar mass of $\sim3-4\times10^{10}\msun$, the mass-size scaling relations steepen with less scatter for both late- and early-types. However, the early-types have a significantly steeper relation than late-types. 

In S03, the mass-size relation for early-types is reported down to a stellar mass of $\sim 10^{10} \msun$. They indicated that faint ellipticals were missed in the analysis due to their type classifications based on concentrations and \ser indices. However, they report tentative evidence that the size-luminosity relation for faint red galaxies flattens at low masses. \cite{graham2008} showed that the size-luminosity relation of elliptical galaxies has a varying slope.  \cite{janz2008} also showed a different size luminosity relation for dwarf and giant early-type galaxies in the Virgo Cluster and illustrated that the relation has little to no dependence on luminosity at the faint end. \cite{bernardi2012}  also pointed out the flattening of the early-type mass-size relation, in agreement with what we see in the right panel of Figure \ref{fig4}. With our low-$z$ sample reaching $10^9 \msun$, the flattening of the relation at low masses is clearly seen.

At the high stellar mass end ($\gtrsim 10^{11} \msun$), we find that the sizes of the early-type galaxies tend to be slightly larger than predicted by the S03 relation. \cite{guo2009} also found a similar trend for early-type central galaxies (dashed-dotted lines in Figure \ref{fig4}). We note that the early-types in our sample are morphologically selected and differ from the early-types in S03 (defined as $n>2.5$). In addition, as discussed earlier, the sizes of high \ser index galaxies maybe underestimated in the NYU-VAGC. The number of early-type galaxies in this stellar mass bin is low and provides only weak constraints. However, \cite{bernardi2012} also show similar behavior at this high-mass end. They note that the increase in the steepness of the mass-size relation for high-mass early-types could be due to brightest cluster galaxies \citep[see also][]{bernardi2009, bernardi2007b}. They also pointed out that the steepness of the relation for early-types changes at these high stellar masses.   

However, it is still not clear how the massive early-type galaxies are connected with low-mass galaxies (i.e.,$\lesssim10^{10}\msun$) and how the curvature of the mass-size relation arises for these galaxies. \cite{graham2008} argued that the curved size-luminosity relation for elliptical galaxies is expected from the assumption of varying profile shapes of these galaxies with luminosity and the fact that they are not distinct types. However, \cite{janz2008} find evidence for the different behavior of faint and bright early-types \citep[see also][]{toloba2012}. \cite{bernardi2012} also pointed out that the curvature of the early-type scaling relations might  arise from the presence of other components (e.g., a disk) with the bulges of these galaxies. On the other hand, the characteristic stellar masses discussed above are  predicted by semi-analytical simulations for spheroids in \cite{shankar2013}. They show that the physical processes behind the evolution of spheroid sizes are different below and above these masses, which might naturally explain the differing relations.

For late-type galaxies, we find that the stellar mass-size relation is mostly consistent among our samples, regardless of the exact definition. These relations are also consistent with the mass-size relation for late-types in S03. However, the mass-size relation for blue galaxies is somewhat offset to larger sizes and steeper than that derived for other ``late-type'' classifications. This is likely a consequence of the strong color-size relation pointed out by \cite{franx2008}, as well as the exclusion of red, edge-on spirals from the blue sample \citep[see, e.g.,][]{Patel2012b}. The size-mass relationship among the ``early-type'' samples appears to be consistent regardless of the exact classification method used (elliptical, red, $n>2.5$, and/or quiescent). 

Finally, we artificially redshifted our sample to resemble $z=1$ galaxies in WFC3 $J_{125}$ band images and tested the robustness of size and structural measurements at high redshifts. We re-measure sizes of galaxies with single \ser profile fitting, a common method in the literature for high redshift galaxies. Our results show that using single \ser profile fitting recovers the sizes of these redshifted galaxies without any systematics. Interestingly, this demonstrates that size measurements at high-\textit{z} are robust, despite the single-\ser models failing for nearby massive early-types. We further verified this with simulations, finding that once the small components of nearby two-component galaxies are smeared out at high $z$, single \ser component fitting can adequately measure structural parameters. Image resolution is thus an important criterion for deciding whether to use single \ser profiles. At a physical resolution $\lesssim1$ kpc, where central bright components are well-resolved, overly simple models like single \ser profiles can introduce biases, and a multi-component or non-parametric method should be used.\\

\section{Summary}
We present the mass-size relation of a sample of nearby galaxies at $z=0.01-0.02$, dividing the sample based on several common classifications. We examined different methods of size measurements in order to quantify the systematics associated with each method. We also artificially redshifted these galaxies to $z=1$ to test potential systematic effects on their size measurements at high redshifts.  From our results, we find that:

\begin{itemize}
\item Nearby early-type galaxies with masses $\gtrsim2\times10^{10}\msun$ are not well fit with single \ser profiles. Two-component fits and non-parametric methods appear to provide less biased measurements. These methods produce effective radii that are smaller than those measured with single \ser fits.
\item The stellar mass-size relations of both late-type and early-type galaxies are steep at high masses ($\sim3-4\times10^{10}\msun$) and flatten at low masses. However, this flattening may be affected by the PSF for quiescent galaxies at very low masses ($\log(\mstar/\msun)< 9.5$).
\item Although single-\ser profile fits can be biased for nearby, well-resolved galaxies, they provide robust sizes at high redshifts. 
\item The stellar mass-size relations of ``spiral'' and ``elliptical'' galaxies are not particularly sensitive to the precise definition of these categories (color, \ser index, morphology, sSFR), with the exception of blue galaxies, which follow a somewhat higher and steeper relation.
\end{itemize}

\begin{acknowledgments}
We thank Rychard Bouwens for providing us with the CANDELS DEEP images. We also thank Jarle Brinchmann, Daniel Szomoru, Roozbeh Davari, and Simone Weinmann for useful discussions. 

Funding for the SDSS and SDSS-II has been provided by the Alfred P. Sloan Foundation, the Participating Institutions, the National Science Foundation, the U.S. Department of Energy, the National Aeronautics and Space Administration, the Japanese Monbukagakusho, the Max Planck Society, and the Higher Education Funding Council for England. The SDSS Web site is http://www.sdss.org/.

The SDSS is managed by the Astrophysical Research Consortium for the Participating Institutions. The Participating Institutions are the American Museum of Natural History, Astrophysical Institute Potsdam, University of Basel, University of Cambridge, Case Western Reserve University, University of Chicago, Drexel University, Fermilab, the Institute for Advanced Study, the Japan Participation Group, Johns Hopkins University, the Joint Institute for Nuclear Astrophysics, the Kavli Institute for Particle Astrophysics and Cosmology, the Korean Scientist Group, the Chinese Academy of Sciences (LAMOST), Los Alamos National Laboratory, the Max-Planck-Institute for Astronomy (MPIA), the Max-Planck-Institute for Astrophysics (MPA), New Mexico State University, Ohio State University, University of Pittsburgh, University of Portsmouth, Princeton University, the United States Naval Observatory, and the University of Washington.
\end{acknowledgments}

\appendix
\section{Appendix A}
\subsection{FAILURE OF SINGLE S\'ERSIC FITTING AT $z=0$}

\begin{figure*}
\centering
\includegraphics[height=3.5 in]{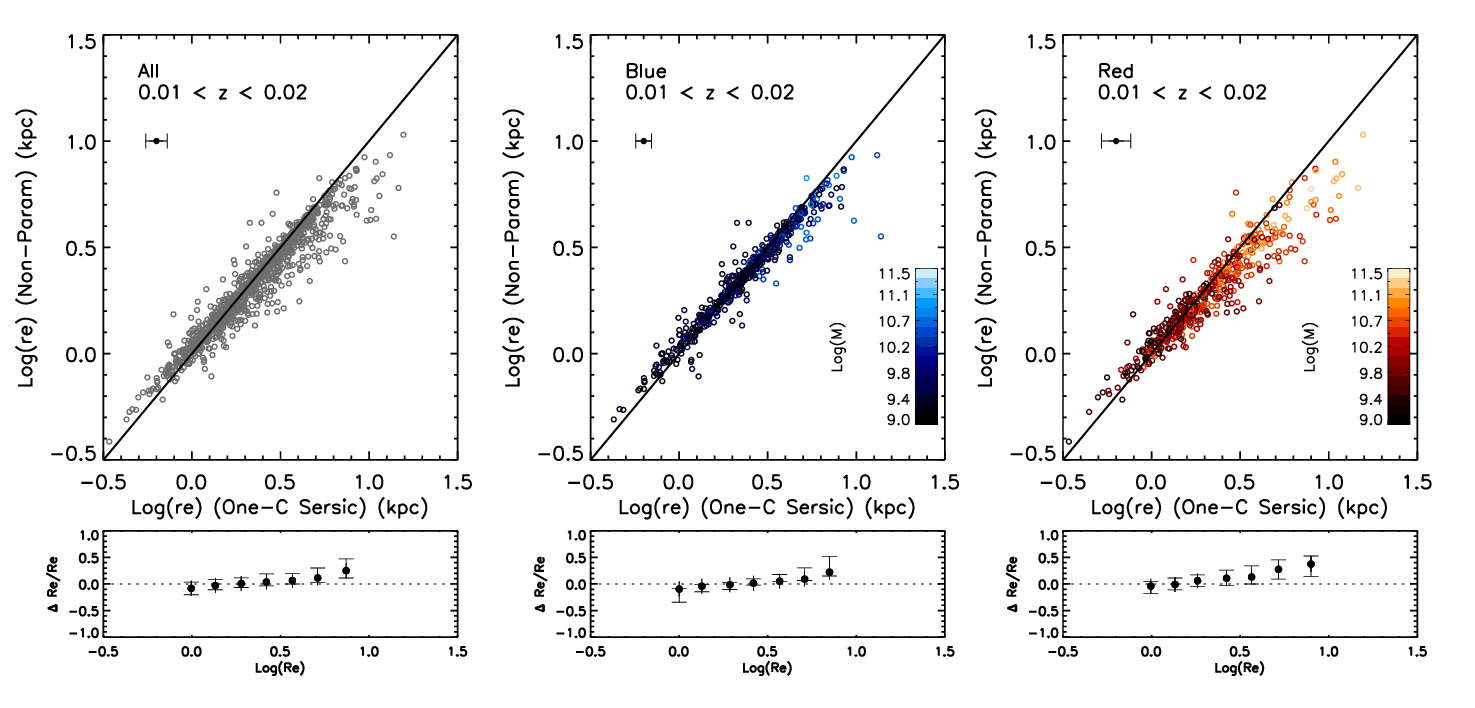}
\caption{Top rows: comparison between sizes of galaxies at $ 0.01 < z < 0.02 $ measured in two different ways, i.e.,  using single component \ser profiles and a non-parametric method, for all galaxies (left panel), blue galaxies (middle panel), and red galaxies (right panel). The bottom panels show the relative differences between sizes of galaxies as a function of their one-component \ser sizes. The systematic differences between sizes of red galaxies increases up to about $40\%$ and sizes based on single \ser profiles are larger that the non-parametric sizes.}
\label{fig10}
\end{figure*}

Surface brightness profiles of galaxies in the local universe rarely conform to simple analytic models \citep[e.g.,][]{allen2006, simard2011}. However, single \ser profile fitting is widely used for measuring structural parameters. In order to test whether using single \ser fitting can bias the sizes of local galaxies with well-resolved profiles, we compare the half-light radii of our sample determined through different methods described in the text. In Figure \ref{fig10}, the sizes of galaxies measured using single \ser profile fitting are compared with their sizes derived from the non-parametric method. In the top-left panel, the comparison is shown for all galaxies and  the relative median differences of sizes as a function of single \ser sizes are illustrated in the bottom-left panel. As can be seen, the median differences are small for small galaxies. However, for large galaxies, the systematic differences reach to $\sim25\%$, i.e., sizes from single \ser profile fitting are systematically larger than sizes from the non-parametric method for these galaxies. To diagnose the systematics, we show the comparison for blue and red galaxies separately in the middle and right panel of Figure \ref{fig10}, respectively. This shows that  the systematics are less than $\sim10\%$ for blue galaxies, except for at the large size end ($\sim20\%$). However, for the red galaxies, the systematic trend is significant and increases toward larger and more massive objects (bottom-right panel).

\begin{figure*}
\centering
\includegraphics[height=3.5 in]{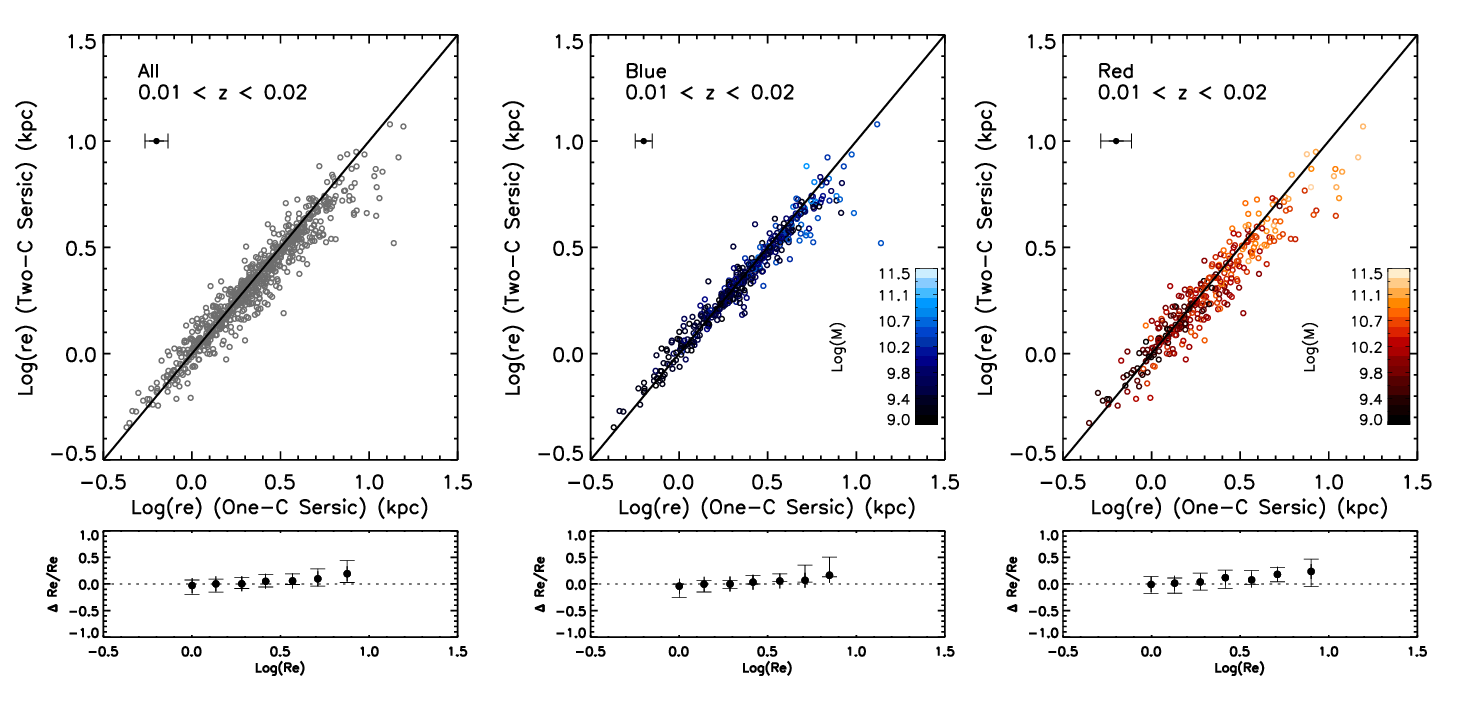}
\caption{Comparison between sizes of galaxies measured using one-component \ser profile fitting and  two-component \ser surface brightness profiles (for $\sim77\%$ of the total sample). As can be seen, galaxies with profiles that could be estimated by two-component \ser profiles have smaller two-component \ser sizes compared with their one-component \ser sizes. }
\label{fig11}
\end{figure*}

We also compare half-light radii from one-component \ser profile fitting and two-component \ser profile fitting in Figure \ref{fig11}. For large and massive galaxies, sizes from single \ser fitting are on average, larger than the sizes from two-component models. Specifically, for the red galaxies (right panels of Figure \ref{fig11}), there is a systematic bias toward larger sizes. 

\begin{figure*}
\centering
\includegraphics[width=\textwidth, height=3in]{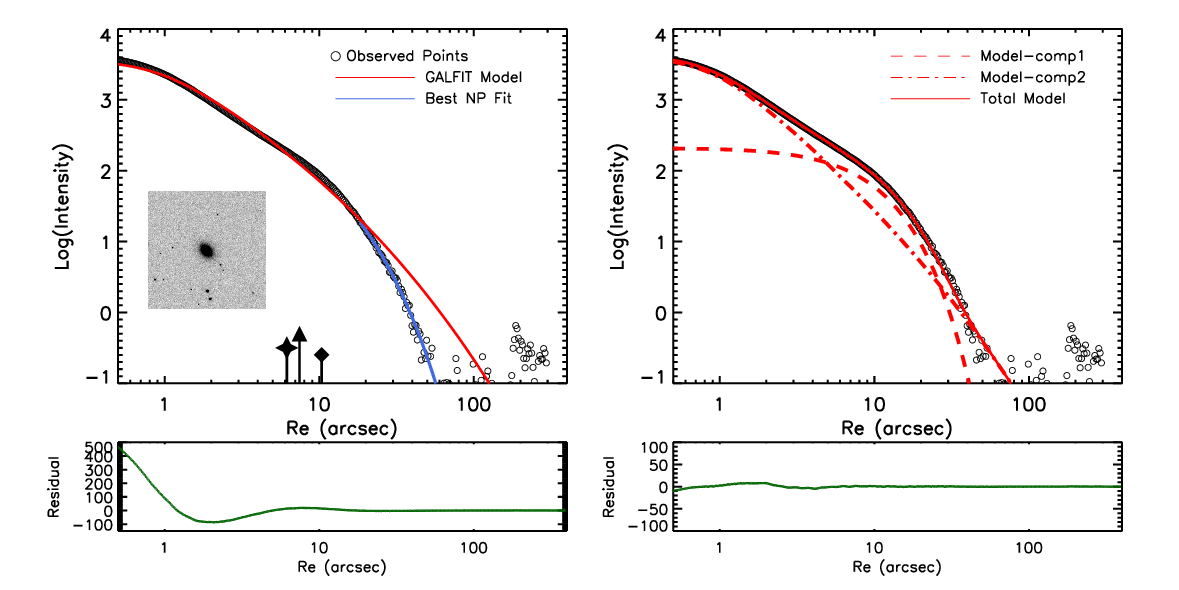}
\caption{Observed profile (black open circles) of a typical early-type galaxy. In the left panel, the red line represents the best-fit one-component \ser profile and the blue line shows the best-fit one-dimensional \ser fit to the outer part of the galaxy ($>petroR90$ for measuring the non-parametric size).  In the right panel, the solid red line represent the total best-fit model from the two-component models. The green lines in the bottom panels show the residuals from the best-fits of one the component \ser profile. The size that is derived using the residual-corrected method \citep{szomoru2010} is shown by a filled star and the non-parametric size is shown by a filled triangle. The filled diamond shows the size of this galaxy using a single-component \ser profile. This plot shows that profiles of galaxies at these very low redshifts might be better explained by two-component profiles. }
\label{fig12}
\end{figure*}

The fact that sizes from single \ser fitting are larger than the sizes from the non-parametric method and the two-component models raises the question of how the half-light radii from one-component \ser profile fitting could have been overestimated. As an example, a typical profile of an early-type galaxy is shown in Figure \ref{fig12}. In the left panel, the observed profile is shown as black circles and the non-parametric fit is overplotted as a blue line. The red line is the best single \ser fit to the galaxy. The single \ser profile to the entire observed profile does not match completely. This can be seen from the extra light in the central regions of the residual profile, which is illustrated in the lower left panel (green line). The half-light size from single \ser fitting is illustrated by the black diamond and is larger than the one derived from the non-parametric method (the black triangle). The size derived using the residual-corrected method \citep{szomoru2010} is also shown as a black star;  this method also produces a smaller size than the single \ser profile fitting. 

The light profile of this galaxy can be described better by adopting two-component \ser profiles. In the upper-right panel of Figure \ref{fig12}, the two-component models and the total model are shown in dashed-dotted and solid red lines, respectively.  The residual profile in the bottom right panel shows that this approach recovers most of the true profile of the galaxy. The half-light size derived from this method for this galaxy is consistent with the non-parametric size and hence is smaller than the value derived from single \ser profile fitting. It is worth testing whether the choice of PSF could introduce uncertainties. For that, we re-measure sizes of this galaxy using a nearby non-saturated star as a PSF. This gives us the same results as before. Therefore, we conclude that choosing the SDSS synthetic PSFs is not the cause of the size biases from single \ser fitting for large galaxies.

This basically shows that if massive galaxies are well-resolved or contain multiple components, structural measurements using a single analytical model could potentially be biased. We also used simulations to show this (see Section 3.3).  \\

\section{Appendix B}

\subsection{HIGH REDSHIFT SIMULATIONS}

In order to check how well we can recover galaxy properties in the $J_{125}$ WFC3 images, we perform simulations by generating $\sim 2300$ synthetic simulated galaxies (assuming a single \ser surface brightness profile) with random properties within the following ranges: $20<J_{125}<26.5$, $0.5< n < 6.5$, and $ 0 < r_e < 15$ kpc$_{z=1}$, convolving with the $J_{125}$-band PSF.  We further add sky background by inserting the simulated galaxy images into the empty regions of the real $J_{125}$-band images and then re-measuring their structural properties with the same procedure that we use for real galaxies. 

\begin{figure*}
\includegraphics[width=\textwidth]{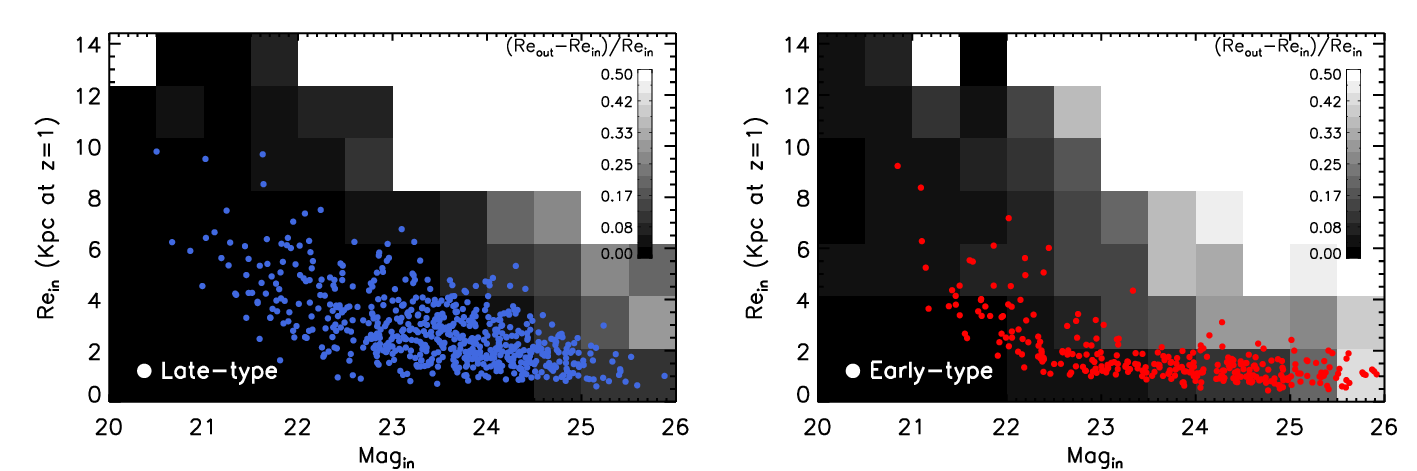}
\caption{Relative differences between input and output sizes of synthetic simulated galaxies (simulation III),  measured in small bins over the (input) size-magnitude plane (left panel: $n < 2.5$; right panel: $ n > 2.5$).  In each bin, the colors correspond to the median relative differences between the recovered and input sizes of the simulated galaxies. The red and blue points represent the artificially redshifted SDSS galaxies (left panel: late-type galaxies; right panel: early-type galaxies) on the size-magnitude plane. This shows that the systematics in the size measurements of our galaxies are very small over their size-magnitude distributions.}

\label{fig13}
\end{figure*}

The results of the simulations are shown in Figure \ref{fig13}. Synthetic simulated galaxies are split into late-type ($n<2.5$; left panel) and early-type ($n>2.5$; right panel), according to their input \ser indices. Then, they are split into small bins over the size-magnitude plane. The relative differences between the input and output sizes (i.e., $\Delta (r_e)/re_{in}$) in each small bin of the size-magnitude distribution are measured and shaded accordingly. Then, we overplot the distribution of artificially redshifted SDSS late-type and early-type galaxies on this size-magnitude plane, shown as blue and red points, respectively. Therefore, systematics in size measurements for each redshifted object can be estimated from this plot.  

As can be seen in Figure \ref{fig13}, for late-type galaxies the systematic differences over the range of artificially redshifted SDSS galaxies are less than a few percent. The systematic differences for early-type galaxies are also very small and only increase at the very faint magnitude end (i.e., $J_{125} > 25$).  In general, comparing the distribution of  artificially redshifted SDSS galaxies with the uncertainties in each bin shows that  the systematics are expected to be very small ($<10\%$ at most) for most of our sample. Therefore, we expect that our size measurement procedure at high-\textit{z} recovers the properties of galaxies without introducing significant systematic biases. 

\begin{figure*}
\centering
\includegraphics[width=6.4in]{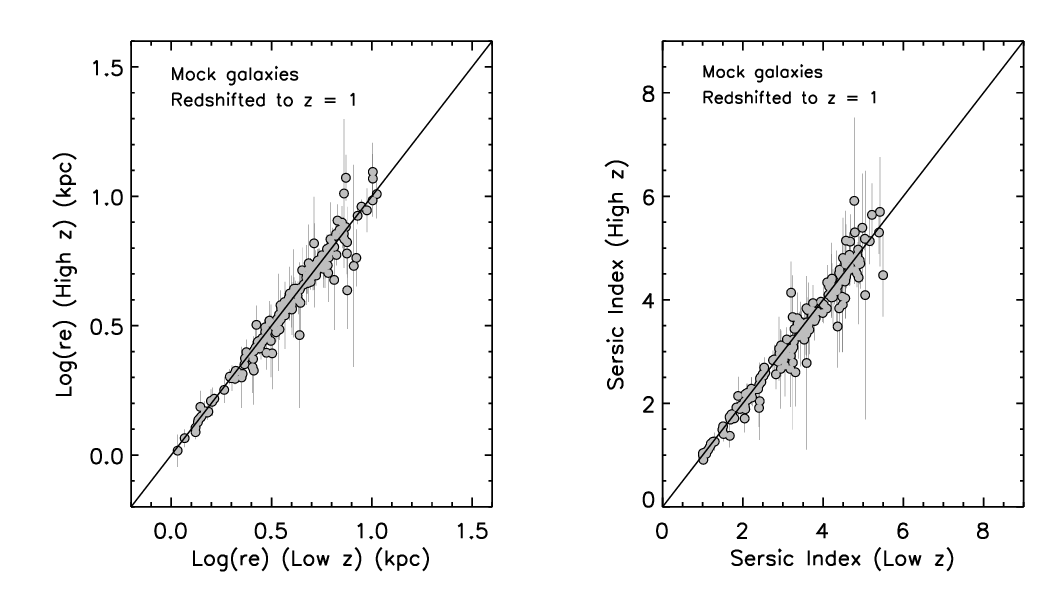}
\caption{Artificial single \ser model SDSS galaxies are created and then redshifted to $z=1$ using our procedure. The left panel shows the comparison between their input sizes (at $z\sim0$) to their sizes after being artificially redshifted. The right panel illustrates the comparison between the input and output \ser indices. The results indicate that our procedure recovers the parameters of \ser model galaxies and that there are no systematics in the measured sizes and \ser indices after redshifting these mock galaxies. }

\label{fig14}
\end{figure*}

The next set of simulations is designed to check our redshifting procedure. For this purpose, we also create two-dimensional single \ser model galaxies with a similar range of properties as nearby SDSS galaxies. We assign them similar redshifts as our SDSS galaxy sample. Then, we use our code to artificially redshift these mock galaxies to $z=1$, insert them into $J_{125}$ WFC3 images, and re-measure their properties using the method described in Section 5.1. The results are illustrated in Figure \ref{fig14}. In the left panel, input sizes before redshifting and output sizes after redshifting are compared; in the right panel, the comparison of input and output \ser indices is shown. The error bars come from the dispersion between different realizations (i.e., using different empty regions). This plot shows that the properties of these single \ser model galaxies can be recovered after redshifting to $z=1$ without any systematics and hence, our redshifting procedure works robustly. 

\begin{figure*}
\centering
\includegraphics[width=\textwidth]{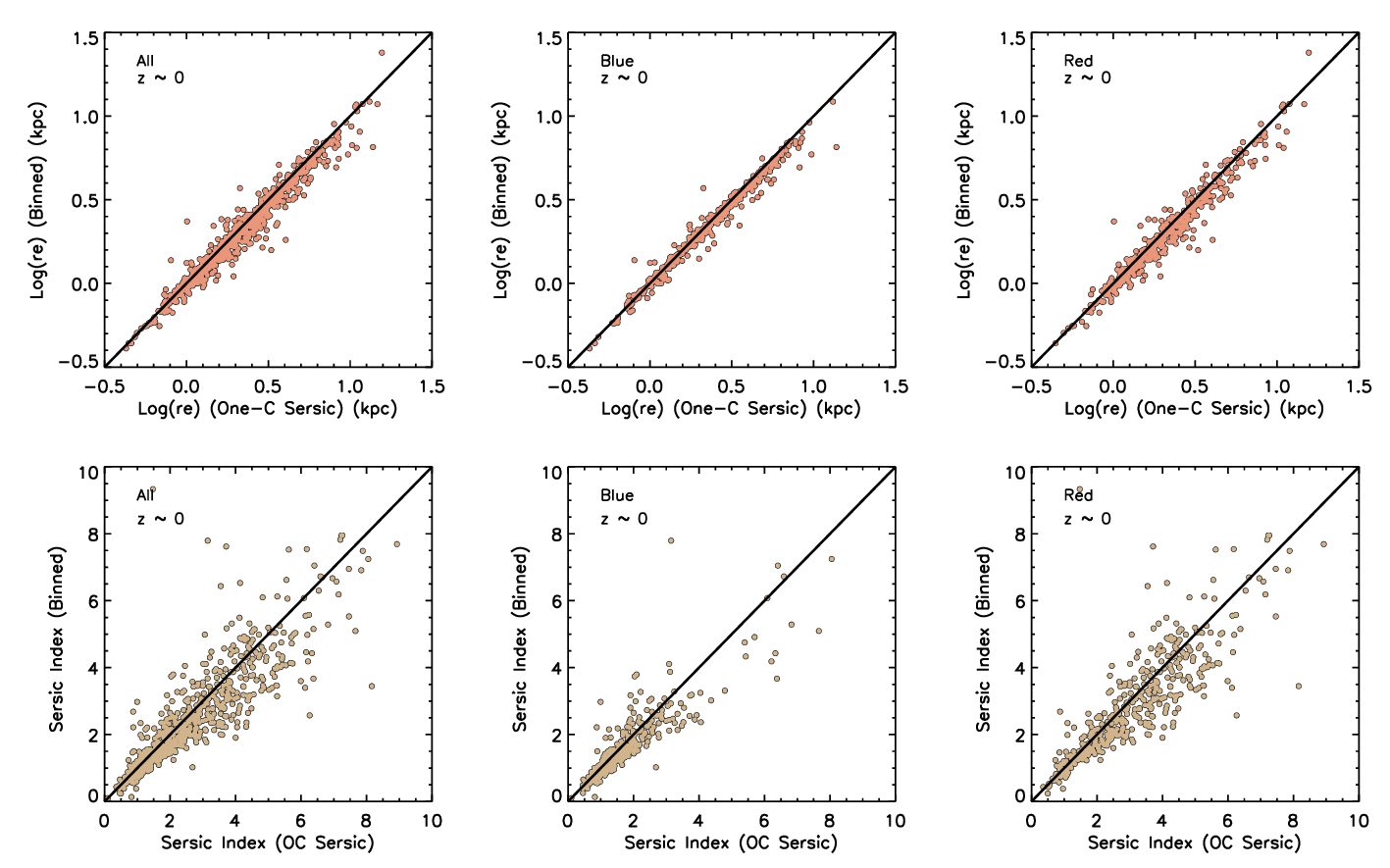}
\caption{Comparison of sizes (top row) and \ser indices (bottom row) of galaxies at $z=0$ derived from single-component \ser fits of SDSS galaxies before (labeled as one-component) and after (labeled as binned; through binning and smoothing) degrading the resolution. This test shows that using single \ser fitting for well-resolved images of nearby galaxies could result in the  overestimate on of parameters such as sizes and \ser indices, compared with lower resolution images.}

\label{fig15}
\end{figure*}

We also discuss in the text that using single \ser profile fitting likely measures the true structural parameters of galaxies at high redshifts. We verify this by simulating double-component galaxies and redshifting them to $z=1$ (simulation (II) and Figure \ref{fig7}).  In addition to these results, in order to test  whether using low resolution images washes out the sub-components and changes the measured structural parameters, we re-measure galaxy sizes at $z=0$ using single \ser profile fitting from their degraded images; i.e., images that are binned (by a factor of four) and Gaussian-smoothed.  Figure \ref{fig15} shows the sizes (top panels) and \ser indices (bottom panels) of these galaxies after smearing. It can be seen that the sizes and \ser indices are smaller, especially for red galaxies after degrading. The results simply illustrate that the bright central parts of galaxies can bias measurements of the structural properties of  galaxies in the nearby universe, when a single \ser model is used. Also, they show that resolution should be taken into account for structural parameter measurements. \\

\begin{thebibliography}{67}
\expandafter\ifx\csname natexlab\endcsname\relax\def\natexlab#1{#1}\fi

\bibitem[{{Abazajian} {et~al.}(2009){Abazajian}, {Adelman-McCarthy},
  {Ag{\"u}eros}, {Allam}, {Allende Prieto}, {An}, {Anderson}, {Anderson},
  {Annis}, {Bahcall}, \& et~al.}]{abazajian2009}
{Abazajian}, K.~N., {et~al.} 2009, \apjs, 182, 543

\bibitem[{{Allen} {et~al.}(2006){Allen}, {Driver}, {Graham}, {Cameron},
  {Liske}, \& {de Propris}}]{allen2006}
{Allen}, P.~D., {Driver}, S.~P., {Graham}, A.~W., {Cameron}, E., {Liske}, J.,
  \& {de Propris}, R. 2006, \mnras, 371, 2

\bibitem[{{Barden} {et~al.}(2008){Barden}, {Jahnke}, \&
  {H{\"a}u{\ss}ler}}]{barden2008}
{Barden}, M., {Jahnke}, K., \& {H{\"a}u{\ss}ler}, B. 2008, \apjs, 175, 105

\bibitem[{{Barden} {et~al.}(2005){Barden}, {Rix}, {Somerville}, {Bell},
  {H{\"a}u{\ss}ler}, {Peng}, {Borch}, {Beckwith}, {Caldwell}, {Heymans},
  {Jahnke}, {Jogee}, {McIntosh}, {Meisenheimer}, {S{\'a}nchez}, {Wisotzki}, \&
  {Wolf}}]{barden2005}
{Barden}, M., {et~al.} 2005, \apj, 635, 959

\bibitem[{{Bernardi}(2009)}]{bernardi2009}
{Bernardi}, M. 2009, \mnras, 395, 1491

\bibitem[{{Bernardi} {et~al.}(2007){Bernardi}, {Hyde}, {Sheth}, {Miller}, \&
  {Nichol}}]{bernardi2007b}
{Bernardi}, M., {Hyde}, J.~B., {Sheth}, R.~K., {Miller}, C.~J., \& {Nichol},
  R.~C. 2007, \aj, 133, 1741

\bibitem[{{Bernardi} {et~al.}(2012){Bernardi}, {Meert}, {Vikram},
  {Huertas-Company}, {Mei}, {Shankar}, \& {Sheth}}]{bernardi2012}
{Bernardi}, M., {Meert}, A., {Vikram}, V., {Huertas-Company}, M., {Mei}, S.,
  {Shankar}, F., \& {Sheth}, R.~K. 2012, arXiv:1211.6122


\bibitem[{{Bertin} \& {Arnouts}(1996)}]{bertin1996}
{Bertin}, E., \& {Arnouts}, S. 1996, \aaps, 117, 393

\bibitem[{{Blanton} {et~al.}(2005){Blanton}, {Schlegel}, {Strauss},
  {Brinkmann}, {Finkbeiner}, {Fukugita}, {Gunn}, {Hogg}, {Ivezi{\'c}}, {Knapp},
  {Lupton}, {Munn}, {Schneider}, {Tegmark}, \& {Zehavi}}]{Blanton2005}
{Blanton}, M.~R., {et~al.} 2005, \aj, 129, 2562

\bibitem[{{Bouwens} {et~al.}(2012){Bouwens}, {Illingworth}, {Oesch}, {Franx},
  {Labb{\'e}}, {Trenti}, {van Dokkum}, {Carollo}, {Gonz{\'a}lez}, {Smit}, \&
  {Magee}}]{bouwens2012}
{Bouwens}, R.~J., {et~al.} 2012, \apj, 754, 83

\bibitem[{{Brinchmann} {et~al.}(2004){Brinchmann}, {Charlot}, {White},
  {Tremonti}, {Kauffmann}, {Heckman}, \& {Brinkmann}}]{brinchmann2004}
{Brinchmann}, J., {Charlot}, S., {White}, S.~D.~M., {Tremonti}, C.,
  {Kauffmann}, G., {Heckman}, T., \& {Brinkmann}, J. 2004, \mnras, 351, 1151

\bibitem[{{Buitrago} {et~al.}(2008){Buitrago}, {Trujillo}, {Conselice},
  {Bouwens}, {Dickinson}, \& {Yan}}]{buitrago2008}
{Buitrago}, F., {Trujillo}, I., {Conselice}, C.~J., {Bouwens}, R.~J.,
  {Dickinson}, M., \& {Yan}, H. 2008, \apjl, 687, L61

\bibitem[{{Cimatti} {et~al.}(2008){Cimatti}, {Cassata}, {Pozzetti}, {Kurk},
  {Mignoli}, {Renzini}, {Daddi}, {Bolzonella}, {Brusa}, {Rodighiero},
  {Dickinson}, {Franceschini}, {Zamorani}, {Berta}, {Rosati}, \&
  {Halliday}}]{cimatti2008}
{Cimatti}, A., {et~al.} 2008, \aap, 482, 21

\bibitem[{{Conselice} {et~al.}(2011){Conselice}, {Bluck}, {Ravindranath},
  {Mortlock}, {Koekemoer}, {Buitrago}, {Gr{\"u}tzbauch}, \&
  {Penny}}]{conselice2011}
{Conselice}, C.~J., {Bluck}, A.~F.~L., {Ravindranath}, S., {Mortlock}, A.,
  {Koekemoer}, A.~M., {Buitrago}, F., {Gr{\"u}tzbauch}, R., \& {Penny}, S.~J.
  2011, \mnras, 417, 2770

\bibitem[{{C{\^o}t{\'e}} {et~al.}(2006){C{\^o}t{\'e}}, {Piatek}, {Ferrarese},
  {Jord{\'a}n}, {Merritt}, {Peng}, {Ha{\c s}egan}, {Blakeslee}, {Mei}, {West},
  {Milosavljevi{\'c}}, \& {Tonry}}]{cote2006}
{C{\^o}t{\'e}}, P., {et~al.} 2006, \apjs, 165, 57

\bibitem[{{Daddi} {et~al.}(2005){Daddi}, {Renzini}, {Pirzkal}, {Cimatti},
  {Malhotra}, {Stiavelli}, {Xu}, {Pasquali}, {Rhoads}, {Brusa}, {di Serego
  Alighieri}, {Ferguson}, {Koekemoer}, {Moustakas}, {Panagia}, \&
  {Windhorst}}]{daddi2005}
{Daddi}, E., {et~al.} 2005, \apj, 626, 680

\bibitem[{{de Vaucouleurs}(1948)}]{devaucouleurs1948}
{de Vaucouleurs}, G. 1948, Annales d'Astrophysique, 11, 247

\bibitem[{{Dutton} {et~al.}(2011){Dutton}, {van den Bosch}, {Faber}, {Simard},
  {Kassin}, {Koo}, {Bundy}, {Huang}, {Weiner}, {Cooper}, {Newman}, {Mozena}, \&
  {Koekemoer}}]{dutton2011}
{Dutton}, A.~A., {et~al.} 2011, \mnras, 410, 1660

\bibitem[{{Ferrarese} {et~al.}(1994){Ferrarese}, {van den Bosch}, {Ford},
  {Jaffe}, \& {O'Connell}}]{ferrarese1994}
{Ferrarese}, L., {van den Bosch}, F.~C., {Ford}, H.~C., {Jaffe}, W., \&
  {O'Connell}, R.~W. 1994, \aj, 108, 1598

\bibitem[{{Franx} {et~al.}(2008){Franx}, {van Dokkum}, {Schreiber}, {Wuyts},
  {Labb{\'e}}, \& {Toft}}]{franx2008}
{Franx}, M., {van Dokkum}, P.~G., {Schreiber}, N.~M.~F., {Wuyts}, S.,
  {Labb{\'e}}, I., \& {Toft}, S. 2008, \apj, 688, 770

\bibitem[{{Giavalisco} {et~al.}(1996){Giavalisco}, {Livio}, {Bohlin},
  {Macchetto}, \& {Stecher}}]{giavalisco1996}
{Giavalisco}, M., {Livio}, M., {Bohlin}, R.~C., {Macchetto}, F.~D., \&
  {Stecher}, T.~P. 1996, \aj, 112, 369

\bibitem[{{Graham} \& {Driver}(2005)}]{graham2005}
{Graham}, A.~W., \& {Driver}, S.~P. 2005, pasa, 22, 118

\bibitem[{{Graham} {et~al.}(2003){Graham}, {Erwin}, {Trujillo}, \& {Asensio
  Ramos}}]{graham2003}
{Graham}, A.~W., {Erwin}, P., {Trujillo}, I., \& {Asensio Ramos}, A. 2003, \aj,
  125, 2951

\bibitem[{{Graham} \& {Worley}(2008)}]{graham2008}
{Graham}, A.~W., \& {Worley}, C.~C. 2008, \mnras, 388, 1708

\bibitem[{{Guo} {et~al.}(2009){Guo}, {McIntosh}, {Mo}, {Katz}, {van den Bosch},
  {Weinberg}, {Weinmann}, {Pasquali}, \& {Yang}}]{guo2009}
{Guo}, Y., {et~al.} 2009, \mnras, 398, 1129

\bibitem[{{Hopkins} {et~al.}(2009{\natexlab{a}}){Hopkins}, {Cox}, {Dutta},
  {Hernquist}, {Kormendy}, \& {Lauer}}]{hopkins2009b}
{Hopkins}, P.~F., {Cox}, T.~J., {Dutta}, S.~N., {Hernquist}, L., {Kormendy},
  J., \& {Lauer}, T.~R. 2009{\natexlab{a}}, \apjs, 181, 135

\bibitem[{{Hopkins} {et~al.}(2009{\natexlab{b}}){Hopkins}, {Lauer}, {Cox},
  {Hernquist}, \& {Kormendy}}]{hopkins2009a}
{Hopkins}, P.~F., {Lauer}, T.~R., {Cox}, T.~J., {Hernquist}, L., \& {Kormendy},
  J. 2009{\natexlab{b}}, \apjs, 181, 486

\bibitem[Huang et al.(2013)]{huang2012} Huang, S., Ho, L.~C., 
Peng, C.~Y., Li, Z.-Y., \& Barth, A.~J.\ 2013, \apj, 766, 47 

\bibitem[{{Janz} \& {Lisker}(2008)}]{janz2008}
{Janz}, J., \& {Lisker}, T. 2008, \apjl, 689, L25

\bibitem[{{Jedrzejewski}(1987)}]{jedrzejewski1987}
{Jedrzejewski}, R.~I. 1987, \mnras, 226, 747

\bibitem[{{Kauffmann} {et~al.}(2003){Kauffmann}, {Heckman}, {White}, {Charlot},
  {Tremonti}, {Peng}, {Seibert}, {Brinkmann}, {Nichol}, {SubbaRao}, \&
  {York}}]{kauffmann2003}
{Kauffmann}, G., {et~al.} 2003, \mnras, 341, 54

\bibitem[{{Kormendy}(1985)}]{kormendy85}
{Kormendy}, J. 1985, \apj, 295, 73

\bibitem[{{Kormendy} {et~al.}(2009){Kormendy}, {Fisher}, {Cornell}, \&
  {Bender}}]{kormendy2009}
{Kormendy}, J., {Fisher}, D.~B., {Cornell}, M.~E., \& {Bender}, R. 2009, \apjs,
  182, 216

\bibitem[{{Labb{\'e}} {et~al.}(2003){Labb{\'e}}, {Rudnick}, {Franx}, {Daddi},
  {van Dokkum}, {F{\"o}rster Schreiber}, {Kuijken}, {Moorwood}, {Rix},
  {R{\"o}ttgering}, {Trujillo}, {van der Wel}, {van der Werf}, \& {van
  Starkenburg}}]{labbe2003}
{Labb{\'e}}, I., {et~al.} 2003, \apjl, 591, L95

\bibitem[{{Lauer} {et~al.}(1995){Lauer}, {Ajhar}, {Byun}, {Dressler}, {Faber},
  {Grillmair}, {Kormendy}, {Richstone}, \& {Tremaine}}]{lauer1995}
{Lauer}, T.~R., {et~al.} 1995, \aj, 110, 2622

\bibitem[{{Lauer} {et~al.}(2007){Lauer}, {Gebhardt}, {Faber}, {Richstone},
  {Tremaine}, {Kormendy}, {Aller}, {Bender}, {Dressler}, {Filippenko}, {Green},
  \& {Ho}}]{lauer2007}
---. 2007, \apj, 664, 226

\bibitem[{{Law} {et~al.}(2012){Law}, {Steidel}, {Shapley}, {Nagy}, {Reddy}, \&
  {Erb}}]{law2011}
{Law}, D.~R., {Steidel}, C.~C., {Shapley}, A.~E., {Nagy}, S.~R., {Reddy},
  N.~A., \& {Erb}, D.~K. 2012, \apj, 745, 85

\bibitem[{{Lintott} {et~al.}(2011){Lintott}, {Schawinski}, {Bamford}, {Slosar},
  {Land}, {Thomas}, {Edmondson}, {Masters}, {Nichol}, {Raddick}, {Szalay},
  {Andreescu}, {Murray}, \& {Vandenberg}}]{lintott2011}
{Lintott}, C., {et~al.} 2011, \mnras, 410, 166

\bibitem[{{Lisker} {et~al.}(2006){Lisker}, {Debattista}, {Ferreras}, \&
  {Erwin}}]{lisker2006}
{Lisker}, T., {Debattista}, V.~P., {Ferreras}, I., \& {Erwin}, P. 2006, \mnras,
  370, 477

\bibitem[{{Mancini} {et~al.}(2010){Mancini}, {Daddi}, {Renzini}, {Salmi},
  {McCracken}, {Cimatti}, {Onodera}, {Salvato}, {Koekemoer}, {Aussel}, {Le
  Floc'h}, {Willott}, \& {Capak}}]{mancini2010}
{Mancini}, C., {et~al.} 2010, \mnras, 401, 933

\bibitem[Meert et al.(2013)]{meert2012} Meert, A., Vikram, V., 
\& Bernardi, M.\ 2013, \mnras, 433, 1344 

\bibitem[{{Mosleh} {et~al.}(2011){Mosleh}, {Williams}, {Franx}, \&
  {Kriek}}]{mosleh2011}
{Mosleh}, M., {Williams}, R.~J., {Franx}, M., \& {Kriek}, M. 2011, \apj, 727, 5

\bibitem[{{Mosleh} {et~al.}(2012){Mosleh}, {Williams}, {Franx}, {Gonzalez},
  {Bouwens}, {Oesch}, {Labbe}, {Illingworth}, \& {Trenti}}]{mosleh2012}
{Mosleh}, M., {et~al.} 2012, \apjl, 756, L12

\bibitem[{{Newman} {et~al.}(2012){Newman}, {Ellis}, {Bundy}, \&
  {Treu}}]{newman2012}
{Newman}, A.~B., {Ellis}, R.~S., {Bundy}, K., \& {Treu}, T. 2012, \apj, 746,
  162

\bibitem[{{Oesch} {et~al.}(2010){Oesch}, {Bouwens}, {Carollo}, {Illingworth},
  {Trenti}, {Stiavelli}, {Magee}, {Labb{\'e}}, \& {Franx}}]{oesch2010b}
{Oesch}, P.~A., {et~al.} 2010, \apjl, 709, L21

\bibitem[{{Patel} {et~al.}(2012{\natexlab{a}}){Patel}, {Holden}, {Kelson},
  {Franx}, {van der Wel}, \& {Illingworth}}]{Patel2012b}
{Patel}, S.~G., {Holden}, B.~P., {Kelson}, D.~D., {Franx}, M., {van der Wel},
  A., \& {Illingworth}, G.~D. 2012{\natexlab{a}}, \apjl, 748, L27

\bibitem[Patel et al.(2013)]{patel2012} Patel, S.~G., van Dokkum, 
P.~G., Franx, M., et al.\ 2013, \apj, 766, 15 


\bibitem[{{Peng} {et~al.}(2010){Peng}, {Ho}, {Impey}, \& {Rix}}]{peng2010}
{Peng}, C.~Y., {Ho}, L.~C., {Impey}, C.~D., \& {Rix}, H.-W. 2010, \aj, 139,
  2097

\bibitem[{{Petty} {et~al.}(2009){Petty}, {de Mello}, {Gallagher}, {Gardner},
  {Lotz}, {Mountain}, \& {Smith}}]{petty2009}
{Petty}, S.~M., {de Mello}, D.~F., {Gallagher}, III, J.~S., {Gardner}, J.~P.,
  {Lotz}, J.~M., {Mountain}, C.~M., \& {Smith}, L.~J. 2009, \aj, 138, 362

\bibitem[{{Salim} {et~al.}(2007){Salim}, {Rich}, {Charlot}, {Brinchmann},
  {Johnson}, {Schiminovich}, {Seibert}, {Mallery}, {Heckman}, {Forster},
  {Friedman}, {Martin}, {Morrissey}, {Neff}, {Small}, {Wyder}, {Bianchi},
  {Donas}, {Lee}, {Madore}, {Milliard}, {Szalay}, {Welsh}, \& {Yi}}]{salim2007}
{Salim}, S., {et~al.} 2007, \apjs, 173, 267

\bibitem[{{S{\'e}rsic}(1963)}]{sersic1963}
{S{\'e}rsic}, J.~L. 1963, Boletin de la Asociacion Argentina de Astronomia La
  Plata Argentina, 6, 41

\bibitem[Sersic(1968)]{sersic1968} Sersic, J.~L.\ 1968, Cordoba, 
Argentina: Observatorio Astronomico, 1968, 

\bibitem[{{Shankar} {et~al.}(2013){Shankar}, {Marulli}, {Bernardi}, {Mei},
  {Meert}, \& {Vikram}}]{shankar2013}
{Shankar}, F., {Marulli}, F., {Bernardi}, M., {Mei}, S., {Meert}, A., \&
  {Vikram}, V. 2013, \mnras, 428, 109

\bibitem[{{Shen} {et~al.}(2003){Shen}, {Mo}, {White}, {Blanton}, {Kauffmann},
  {Voges}, {Brinkmann}, \& {Csabai}}]{shen2003}
{Shen}, S., {Mo}, H.~J., {White}, S.~D.~M., {Blanton}, M.~R., {Kauffmann}, G.,
  {Voges}, W., {Brinkmann}, J., \& {Csabai}, I. 2003, \mnras, 343, 978

\bibitem[{{Simard} {et~al.}(2011){Simard}, {Mendel}, {Patton}, {Ellison}, \&
  {McConnachie}}]{simard2011}
{Simard}, L., {Mendel}, J.~T., {Patton}, D.~R., {Ellison}, S.~L., \&
  {McConnachie}, A.~W. 2011, \apjs, 196, 11

\bibitem[{{Szomoru} {et~al.}(2012){Szomoru}, {Franx}, \& {van
  Dokkum}}]{szomoru2012}
{Szomoru}, D., {Franx}, M., \& {van Dokkum}, P.~G. 2012, \apj, 749, 121

\bibitem[{{Szomoru} {et~al.}(2010){Szomoru}, {Franx}, {van Dokkum}, {Trenti},
  {Illingworth}, {Labb{\'e}}, {Bouwens}, {Oesch}, \& {Carollo}}]{szomoru2010}
{Szomoru}, D., {et~al.} 2010, \apjl, 714, L244

\bibitem[{{Taylor} {et~al.}(2010){Taylor}, {Franx}, {Glazebrook}, {Brinchmann},
  {van der Wel}, \& {van Dokkum}}]{Taylor2010}
{Taylor}, E.~N., {Franx}, M., {Glazebrook}, K., {Brinchmann}, J., {van der
  Wel}, A., \& {van Dokkum}, P.~G. 2010, \apj, 720, 723

\bibitem[{{Toloba} {et~al.}(2012){Toloba}, {Boselli}, {Peletier},
  {Falc{\'o}n-Barroso}, {van de Ven}, \& {Gorgas}}]{toloba2012}
{Toloba}, E., {Boselli}, A., {Peletier}, R.~F., {Falc{\'o}n-Barroso}, J., {van
  de Ven}, G., \& {Gorgas}, J. 2012, \aap, 548, A78

\bibitem[{{Trujillo} {et~al.}(2006){Trujillo}, {F{\"o}rster Schreiber},
  {Rudnick}, {Barden}, {Franx}, {Rix}, {Caldwell}, {McIntosh}, {Toft},
  {H{\"a}ussler}, {Zirm}, {van Dokkum}, {Labb{\'e}}, {Moorwood},
  {R{\"o}ttgering}, {van der Wel}, {van der Werf}, \& {van
  Starkenburg}}]{trujillo2006a}
{Trujillo}, I., {et~al.} 2006, \apj, 650, 18

\bibitem[{{Turner} {et~al.}(2012){Turner}, {C{\^o}t{\'e}}, {Ferrarese},
  {Jord{\'a}n}, {Blakeslee}, {Mei}, {Peng}, \& {West}}]{turner2012}
{Turner}, M.~L., {C{\^o}t{\'e}}, P., {Ferrarese}, L., {Jord{\'a}n}, A.,
  {Blakeslee}, J.~P., {Mei}, S., {Peng}, E.~W., \& {West}, M.~J. 2012, \apjs,
  203, 5

\bibitem[van de Sande et al.(2013)]{sande2012} van de Sande, J., 
Kriek, M., Franx, M., et al.\ 2013, \apj, 771, 85 

\bibitem[{{van den Bergh} {et~al.}(2002){van den Bergh}, {Abraham}, {Whyte},
  {Merrifield}, {Eskridge}, {Frogel}, \& {Pogge}}]{vanderbergh2002}
{van den Bergh}, S., {Abraham}, R.~G., {Whyte}, L.~F., {Merrifield}, M.~R.,
  {Eskridge}, P.~B., {Frogel}, J.~A., \& {Pogge}, R. 2002, \aj, 123, 2913

\bibitem[{{van der Wel} {et~al.}(2008){van der Wel}, {Holden}, {Zirm}, {Franx},
  {Rettura}, {Illingworth}, \& {Ford}}]{vanderwel2008}
{van der Wel}, A., {Holden}, B.~P., {Zirm}, A.~W., {Franx}, M., {Rettura}, A.,
  {Illingworth}, G.~D., \& {Ford}, H.~C. 2008, \apj, 688, 48

\bibitem[{{van Dokkum} {et~al.}(2008){van Dokkum}, {Franx}, {Kriek}, {Holden},
  {Illingworth}, {Magee}, {Bouwens}, {Marchesini}, {Quadri}, {Rudnick},
  {Taylor}, \& {Toft}}]{vandokkum2008}
{van Dokkum}, P.~G., {et~al.} 2008, \apjl, 677, L5

\bibitem[{{Weinzirl} {et~al.}(2011){Weinzirl}, {Jogee}, {Conselice},
  {Papovich}, {Chary}, {Bluck}, {Gr{\"u}tzbauch}, {Buitrago}, {Mobasher},
  {Lucas}, {Dickinson}, \& {Bauer}}]{weinzirl2011}
{Weinzirl}, T., {et~al.} 2011, \apj, 743, 87

\bibitem[{{Williams} {et~al.}(2010){Williams}, {Quadri}, {Franx}, {van Dokkum},
  {Toft}, {Kriek}, \& {Labb{\'e}}}]{williams2009}
{Williams}, R.~J., {Quadri}, R.~F., {Franx}, M., {van Dokkum}, P., {Toft}, S.,
  {Kriek}, M., \& {Labb{\'e}}, I. 2010, \apj, 713, 738

\end{thebibliography}
\end{document}